\documentclass[11pt]{article}
\usepackage{setspace} 
\usepackage[utf8]{inputenc}
\usepackage[english]{babel}
\usepackage{graphicx}
\usepackage{enumerate}
\usepackage{amssymb}
\usepackage{amsmath}
\usepackage{amsthm}

\usepackage{caption}
\usepackage{subcaption}
\usepackage{ragged2e}
\usepackage{lscape}
\usepackage{hyperref}
\usepackage{multirow}
\usepackage[margin=2.5cm]{geometry}
\usepackage[capposition=bottom]{floatrow}
\usepackage{comment}

\newcommand{\fixme}[1]{}
\newcommand{\workinprogress}[1]{}

\usepackage{tikz}
\usepackage[
backend=biber,
style=authoryear,
natbib=true]{biblatex}
\AtEveryBibitem{\clearfield{issn}}
\AtEveryBibitem{\clearfield{doi}}
\AtEveryBibitem{\clearfield{url}}
\AtEveryBibitem{\clearfield{month}}
\newtheorem{theorem}{Theorem}
\newtheorem{lemma}{Lemma}
\newtheorem{proposition}{Proposition}

\newtheorem{definition}{Definition}
\newtheorem{example}{Example}
\newtheorem{hypothesis}{Hypothesis}

\title{Voluntary Information Disclosure in Centralized Matching:\\ Efficiency Gains and Strategic Properties\footnote{We are thankful to participants at the Easter Workshop in School Choice at Queen's University Belfast as well as Egor Starkov, Nick Vikander and Mikkel H. Gandil for helpful comments. We are grateful for the funding from Economic Policy Research Network. Any remaining errors are our own.}}

\author{Andreas Bjerre-Nielsen\footnote{University of Copenhagen, email: anbn@econ.ku.dk} \and Emil Chrisander\footnote{University of Copenhagen, email: emil.chrisander@econ.ku.dk}}
\date{This version: \today. \\First draft: June 14, 2021}

\begin{document}

\maketitle
\begin{abstract}
Information frictions can harm the welfare of participants in two-sided matching markets. Consider a centralized admission, where colleges cannot observe students' preparedness for success in a particular major or degree program. Colleges choose between using simple, cheap admission criteria, e.g., high school grades as a proxy for preparedness, or screening all applications, which is time-consuming for both students and colleges. To address issues of fairness and welfare, we introduce two novel mechanisms that allow students to disclose private information voluntarily and thus only require partial screening. The mechanisms are based on Deferred Acceptance and preserve its core strategic properties of credible preference revelation, particularly ordinal strategy-proofness. In addition, we demonstrate conditions for which cardinal welfare improves for market participants compared to not screening. Intuitively, students and colleges benefit from voluntary information disclosure if public information about students correlates weakly with students' private information and the cost of processing disclosed information is sufficiently low. Finally, we present empirical evidence from the Danish higher education system that supports critical features of our model. Our work has policy implications for the mechanism design of large two-sided markets where information frictions are inherent. \end{abstract}

Keywords: market design, school choice, college admission, information frictions, information disclosure

\section{Introduction}

A core challenge in the design of matching mechanisms is to facilitate the disclosure of private information to overcome harmful information frictions.
Deferred Acceptance is a matching mechanism for admission to schools, colleges and in other two-sided matching markets that is increasingly adopted worldwide. It provides a solution to the fundamental problem of not knowing applicants' preferences. 
The mechanism collects credible information about applicant preferences by incentivizing applicants to report their preferences truthfully \citep{Roth1982TheIncentives}. 
However, this mechanism does not solve other dimensions of information frictions, e.g., a low standardized test may deprive otherwise talented students of a second chance of admission to a college. 
Such lack of information about college applicants' academic ability and preparedness may impact overall match efficiency for both applicants and colleges.

Colleges have a complete ranking of all potential applicants in the standard matching framework. Thus, policymakers face a choice between screening all applicants or none to overcome information frictions about applicants' preparedness.
Without screening applicants, colleges prioritize applicants according to an eligibility score, usually computed from standardized tests or prior grade point averages. This approach can be efficient if the eligibility score is a reliable measure of applicants' preparedness as it requires little or no additional screening. Conversely, colleges may rely on additional screening of every applicant if the eligibility score is an unreliable measure of preparedness. Although exhaustive screening may result in a more efficient matching outcome, it is time-consuming for both applicants and colleges \citep{He2019ApplicationMarkets}. Consequently, a natural candidate for efficiency is a matching mechanism that screens only a subset of the applicants. Despite this, no existing work examines such differential screening mechanisms that trade off the efficiency gains of improved matching quality with the increased costs from time spent in making applications and screening potential students.

In this paper, we explore a simplified model of college admission where colleges preferences depend on whether they have screened a given student's application. Students, who are on the demand side of the two-sided matching market, have private information about both their preferences and the payoff to colleges, which represents the ability and preparedness of applicants. We explore how the private information about the payoff to colleges can be voluntarily disclosed to the supply side through a costly application and screening process. Our framework of simplified college admission system combines elements from frameworks used for school choice and college admission. We expand standard school choice by incorporating college payoff from admitting students, which corresponds to incentives from students paying tuition or colleges being paid for student graduation. To make our model tractable, we assume that colleges comply with admission rules and do not engage in strategic behavior and thus always report their priorities truthfully. 

College applicants, who we refer to as students irrespective of whether or not they are admitted, submit a rank-ordered list (ROL) containing colleges they prefer. To assign students to colleges in this setting, we introduce two novel matching mechanisms that allow for information disclosure. 
The first mechanism, Deferred Acceptance with Voluntary Information Quotas (henceforth, DAVID-Q), generalizes the standard Deferred Acceptance by reserving parts of each college's available places for admission to students who voluntarily disclosure information. 
The mechanism allocates the remaining seats according to an eligibility score, which does not require any screening. 
Subsequently, the mechanism splits colleges into subordinate units according to quotas for voluntary information disclosure. It then applies the Student Proposing Deferred Acceptance (henceforth, SPDA) based on the submitted preferences, dual rankings, and quotas to produce a matching. 
The use of quotas with different criteria is similar to the introduction of quotas in controlled school choice with the difference that all students can apply for admission through each quota \citep{Abdulkadiroglu2003SchoolApproach}.

The second mechanism, Deferred Acceptance with Voluntary Information Disclosure (henceforth, DAVID-U), allows students to send additional information about themselves.
Initially, colleges prioritize students according to students' eligibility scores. 
However, eligibility scores are not perfect proxies for the value of students to colleges. 
As a result, colleges update their ranking of students based on the additional information they receive. Using the updated rankings of students and submitted preferences, SPDA produces a matching. 

We show how both mechanisms inherit strategic properties from the standard Deferred Acceptance (henceforth, DA) that runs SPDA with priorities from eligibility score.
First, they satisfy ordinal strategy-proofness for students, implying that they can do no better than to submit their true preference order of colleges. 
Second, they both satisfy justified envy, but the property only holds for students admitted under the same admission criterion for the quota-based mechanism. 
In addition, we introduce a new measure of stability that requires no voluntary information even after voluntary information disclosure. 
We show that DAVID-U satisfies this measure.
While the quota-based mechanism is weaker in terms of justified envy, it protects students because they cannot harm their admission prospects by disclosing information, which is not guaranteed in DAVID-U. 

To facilitate analyses of how voluntary information disclosure affects matching and welfare, we show when equilibria are unique. We demonstrate that the equilibria of the preference revelation games of the mechanisms are unique in large economies under mild conditions of smoothness or regularity in the distribution of types. Moreover, we show that, for a specific share of seats used for disclosure, the equilibria of the two mechanisms coincide. We summarize our results on strategic properties in Table~\ref{tab:overview_properties}.

\begin{table}[]
    \centering
    \begin{tabular}{r|c|c|c}
    \textit{Mechanism property} & DA & DAVID-Q & DAVID-U\\
    \hline
    Ordinal strategy-proofness  & yes & yes & yes \\
    Information disclosure is possible & no & yes & yes \\
    Information disclosure is safe & - & yes & no \\
    Justified envy & college-level & quota-level & - \\
    Justified envy with disclosure & - & - & college-level \\
    Unique equilibrium in large economy  & yes & yes & yes \\
    \end{tabular}
    \caption{Overview of theoretical properties of the matching mechanisms. See the text for description of the three mechanisms and their properties.}
    \label{tab:overview_properties}
\end{table}

We examine the welfare implications of using voluntary information disclosure compared to not screening by using rich simulations of student behavior. In general, we demonstrate that both students and colleges are better off on average with voluntary screening if: i) public information about students correlates weakly with students' private information about their preparedness for a given program;  ii) application and screening costs are sufficiently low; and, iii) students' preparedness for a program correlate strongly with their preferences. Naturally, the first two conditions allow colleges to gain from screening students. The final condition entails that students who are better prepared for a college program also prefer admission there. Consequently, this condition allows students to signal their underlying preferences through information disclosure. Such signaling can help students obtain a better match by overcoming the issue in regular DA of only signalling ordinal preferences \citep{Abdulkadiroglu2009Abdulkadiroglu2009.pdf}.

Finally, we assess two core theoretical predictions in our model by investigating the system of admission to higher education in Denmark and Sweden, which use a quota-based voluntary information mechanism \citep{BUVM2021uddannelsesguiden,UH2021AlternativtUrval}. We focus exclusively on the Danish context as it aligns closely with our theoretical model. 
Following core theoretical predictions, our empirical analysis reveals that costly screening helps colleges predict students' latent ability. Moreover, students who are likely to be rejected using the eligibility score are more likely to voluntarily disclose information. This implies lower total application and screening costs in DAVID-Q compared to a mechanism with mandatory screening of every student.

To illustrate how our framework is set up, we provide Example~\ref{example:introdutory} below. The example starts with the standard model of matching without screening and shows how the introduction of voluntary screening can positively affect welfare for students and colleges.

\begin{example} \label{example:introdutory} 
Consider an economy consisting of two students, Ann and Bob, and two colleges, College A and B. Utility, college payoff, and eligibility score are distributed as follows: 
\begin{table}[H]
\centering
\begin{tabular}{ccc}
\multicolumn{3}{c}{Student utility}                                                   \\ \hline
\multicolumn{1}{|c|}{}          & \multicolumn{1}{c|}{Ann} & \multicolumn{1}{c|}{Bob} \\ \hline
\multicolumn{1}{|c|}{College A} & \multicolumn{1}{c|}{4}   & \multicolumn{1}{c|}{3}   \\ \hline
\multicolumn{1}{|c|}{College B} & \multicolumn{1}{c|}{1}   & \multicolumn{1}{c|}{2}   \\ \hline
\end{tabular}
\hfill
\begin{tabular}{ccc}
\multicolumn{3}{c}{College payoff}                                                   \\ \hline
\multicolumn{1}{|c|}{}          & \multicolumn{1}{c|}{Ann} & \multicolumn{1}{c|}{Bob} \\ \hline
\multicolumn{1}{|c|}{College A} & \multicolumn{1}{c|}{3}   & \multicolumn{1}{c|}{2}   \\ \hline
\multicolumn{1}{|c|}{College B} & \multicolumn{1}{c|}{2}   & \multicolumn{1}{c|}{3}   \\ \hline
\end{tabular}
\hfill
\begin{tabular}{ccc}
\multicolumn{3}{c}{Eligibility score}                                                   \\ \hline
\multicolumn{1}{|c|}{}          & \multicolumn{1}{c|}{Ann} & \multicolumn{1}{c|}{Bob} \\ \hline
\multicolumn{1}{|c|}{College A} & \multicolumn{1}{c|}{2}   & \multicolumn{1}{c|}{3}   \\ \hline
\multicolumn{1}{|c|}{College B} & \multicolumn{1}{c|}{2}   & \multicolumn{1}{c|}{3}   \\ \hline
\end{tabular}
\label{tab:example_1}
\end{table}

We now consider three scenarios. In the first scenario there is no screening, which corresponds to standard Deferred Acceptance. The students report preferences truthfully and College A admitting Bob, due to his higher eligibility score, and College B admitting Ann. The total match value under DA is 4 to both students and colleges. 

In the second scenario we allow for information disclosure such that Ann and Bob can send an application to one or both colleges. If a college screens a student's information, it knows the true payoff. We assume that costs of both application and screening equal 1 each. Under these circumstances, Ann may gain by disclosing her value to College A. Such disclosure would result in a reversed match of College A admitting Ann and College B admitting Bob, which increases the total match value to 5 for both students and colleges after costs of application and screening. 

In the final scenario, colleges screen every student. Therefore, the equilibrium would be the same as in the second scenario. However, Bob would need to make a costly application to College B, which would lower the total value to students and the payoff to colleges to 4 each.
\end{example}

The implementation of voluntary information disclosure could have welfare benefits both compared with an admission system that uses a single eligibility score from standardized tests or high school GPAs, e.g., in China \citep{chen2017chinese}, or systems with full screening of college applicants, as in the US. Moreover, although our focus is on students and colleges, it may be applied more broadly in other contexts with similar information frictions where Deferred Acceptance is employed, e.g., in dating, refugee allocation etc.

Our work is related to several existing strands of literature on matching. The idea of signalling interest in a given study within a mechanism was introduced in the framework of \citet{Abdulkadiroglu2015ExpandingChoice}. We expand this into a more general setting that allows for voluntary screening and show that the mechanism has some of the same core strategic properties.

A strand of existing work examines strategic behavior and efficiency in matching without signalling. The pioneering paper of \citet{Gale1962CollegeMarriage} shows that DA leads to a stable matching and that the order of proposals matters for which stable matching is implemented. \citet{Dubins1981MachiavelliAlgorithm,Roth1982TheIncentives} show that students have no incentives to lie about their underlying preferences in DA; thus, the mechanism is weakly strategy-proof. \citet{Haeringer2009ConstrainedChoice,Fack2019BeyondAdmissions} show that DA is no longer strategy-proof if students incur an application cost.
\citet{Calsamiglia2010ConstrainedStudy} confirm this theory in lab experiments and \citet{He2019ApplicationMarkets} show it in a field experiment of admission to graduate programs.
\citet{He2019ApplicationMarkets} find that while the higher application cost  reduces the number of applications and thus lowers screening costs, the value of the matching outcome is almost unaffected.
Furthermore, existing work has examined whether or not centralized admission yields higher welfare under application and screening costs \citep{Hafalir2018CollegeDecentralized,grenet2022preference}. 
Despite also having application costs, our framework of voluntary screening preserves ordinal strategy-proofness, as the strategic behavior is isolated to the decision of whether or not to disclose information to specific colleges. This is similar to \citet{Abdulkadiroglu2015ExpandingChoice} who show the same property under preference signalling.

Two existing studies also examine partial or differential screening. \citet{Carvalho2019CollegeAnalysis} examines admission with differential screening where the first screening stage filters out students with insufficient qualifications and subsequently all qualified applicants are screened exhaustively. \citet{Lee2017InterviewingMarkets} examine matching when workers and employers must engage in a screening process and then subsequently match where only a subset of potential matches can be assessed. 

In our model, congestion is mitigated by allowing for voluntary disclosure, which acts as a costly signalling. The notion of costly signalling in matching goes back to the classic labor market model in \citet{Spence1973JobSignaling} where workers can signal their productivity. A generalization of this decentralized matching model is provided in \citet{hoppe2009theory} with costly signals on both sides of the market. Another idea is to signal preferences using a limited number of costless signals.  \citet{lee2007signalling,Coles2013PreferenceMarkets} show that signalling of preferences can help overcome congestion in labor markets with screening. The idea of costless signals is evaluated experimentally using a centralized matching mechanism by \citet{Lee2014ProposeMarkets} who find that allowing for signals  increase the likelihood of matching. \citet{Abdulkadiroglu2015ExpandingChoice} also show theoretically that allowing signals that are scarce and costless can overcome DA's inability to convey cardinal signals about preferences. Our framework relies on costly applications that are simultaneously signals of underlying skill but also reflect underlying cardinal preferences. 

Our work is also related to the literature on large matching markets. Uniqueness of matchings is found both by \citet{Abdulkadiroglu2015ExpandingChoice} who investigates DA with tie breaks and \citet{Azevedo2016AMarkets}, who more generally examine stable matchings.  

\section{Model framework}

We begin with outlining our framework of school choice where students can voluntarily disclose information about themselves in a finite economy setting. We consider an economy with a finite set $\mathcal{C}=\{1,2,..,C\}$ of colleges and a finite set $\mathcal{I}=\{1,2,..,I\}$ of students. Colleges have a finite number of seats denoted by the vector $S=\{S_1,S_2,..,S_C\}$. Let $\emptyset$ denote a null set, meaning the outcome of not being assigned to any college. A matching $\mu$ is a mapping $\mu : \mathcal{I} \rightarrow \mathcal{C} \cup \emptyset$ such that $|\mu^{-1}(c)|\le S_c$ for each $c \in\mathcal{C}$; that is, $\mu$ is a many-to-one matching with the property that the number of students assigned to a college does not exceed its capacity.

When the pair of a student $i$ and a college $c$ are matched, the student receives von Neumann-Morgenstern utility (vNM) $v_{i,c}\in \mathbb{R}$ and the college receives payoff $w_{i,c} \in \mathbb{R}$. We summarize students' types by  $\theta_i=(v^\theta_i,e^\theta_i,w^\theta_i)$. The eligibility score is an estimate of the true payoff, i.e., $E[w_{i,c}|e_{i,c}]=e_{i,c}$, which we assume is unbiased. 
We assume that each student's ordinal preferences are strict over colleges $\mathcal{C}$. We also assume that each college in $\mathcal{C}$ has a strict priority ordering over students, regardless of whether they are ranked using value or expected value from an eligibility score. In addition, we assume that the null assignment provides a utility of zero to students  and empty seats give zero payoff to colleges. Our assumptions about utility imply that colleges have responsive preferences, which entails that a college's payoff from admitting a student is independent of other students they admit. 

Student types are realized random variables that follow a joint probability function, $\eta$, over the type space $\Theta$. We assume that $\eta$ is atomless, and thus, the likelihood of drawing students who are indifferent about any two colleges is zero. In terms of information, we assume that the type of every student, $\theta$, including the private information about their value is common knowledge across students. 
Colleges do not observe the true payoff until after being matched, but instead observe an eligibility score $e_{i,c}$, which can be used to compute the expected value $E[w_{i,c}|e_{i,c}]$. 
We note that the special case in our model, where the payoff and eligibility score are perfectly correlated, corresponds to the standard model.

\paragraph{Framework with multiple admission criteria}
We now exhibit a different model that allows colleges to admit using multiple criteria for admission. For each criterion $a$ we associate a part of the college with it;  we call this a sub-college and denote this as $c^a$. The sub-college has an associated quota of $S_c^a$, which satisfies that $\sum_{a}S_c^a=S_c$. In this context, each student $i$ also submits a ROL of sub-colleges, $ROL^{sub}_i$. 

We let a matching between students and sub-colleges be denoted by $\mu^{sub}$. Clearly, such a matching have a corresponding matching between colleges and students where students assigned to sub-college $c^a$ are assigned to its associated college, $c$.

\section{Matching mechanisms and strategic properties}

A mechanism designates a message space for each student as well as an assignment as a function of messages and other parameters of the model. Such parameters include college priorities and capacities. We will consider mechanisms that use students' ordinal preferences and potentially some other auxiliary messages as their message space. We now outline and describe the three different matching mechanisms that we analyze. 

We begin with outlining the fundamental assignment procedure, Student-Proposing Deferred Acceptance (\textbf{SPDA}). The procedure requires that students submit a ROL of colleges and that colleges have a complete ranking of students which we call the priority score. The matching proceeds in iterative rounds characterized as follows:
\begin{itemize}
    \item \textit{Round 1:} Every student applies to her most preferred college. Each college rejects students in excess of its capacity that have the lowest priority score.
    \item \textit{Round k:} Every student who is rejected in step (\textit{k}-1) applies to the next choice on her list. Each college, pooling together new students and those it holds from step (\textit{k}-1), rejects the students in excess of its capacity with lowest priority. The matching finalizes once every student is either matched with a college or has exhausted her list of ranked colleges.
\end{itemize}

\paragraph{DA: Admission by eligibility score}
We let DA refer to the use of SPDA where priority scores equal the eligibility scores of students.

\paragraph{DAVID-Q: Information disclosure with sub-quotas}

We now outline a variation in DA using the framework of multiple admission criteria with specific quotas.
The DAVID-Q mechanism runs DA at the sub-college level with the following two admission criteria: regular admission to sub-college, $c^r$, where eligibility score is used to prioritize students, and; a disclosure sub-college, $c^{vid}$, where students are ranked according to expected payoffs, after information disclosure. We say that college $c$ is targeted with information disclosure by $i$ if $c^{vid}\in ROL^{sub}_i$. Students incur a  non-negative application cost of $\gamma_{\mathcal{I}}$ for each target college. Colleges screen students who have applied to them using information disclosure and incur a non-negative screening cost of $\gamma_{\mathcal{C}}$ each time it is targeted.  We call the disclosure share $\delta_c$, where it holds that $S_{c}^{vid}=\delta_c S_c$ and $S_c^r=(1-\delta_c) S_c$.

\paragraph{DAVID-U: Information disclosure as auxiliary message}
In DAVID-U the students are allowed to submit informative messages to specific colleges. DAVID-U proceeds in the following three steps:
\begin{enumerate}[Step 1:]
    \item Every student $i$ submits their rank-ordered list, denoted by $ROL_i$. In addition, the student decides, which colleges to \textit{target} with information disclosure. The set of targeted colleges for each student is denoted by the set $x_i$. If college $c$ is targeted by student $i$, the college sees the exact payoff from admitting the student. For each message that the student sends, she incurs a non-negative application cost of $\gamma_{\mathcal{I}}$. Colleges are assumed to commit to screen every message sent by students and incur non-negative screening costs of $\gamma_{\mathcal{C}}$ for each message. 
    \item Colleges’ priorities over students equal the expected payoff conditional on available information, i.e., eligibility scores and potential messages. Therefore, if a college receives a message from a student then the college's priority equals the student's payoff, otherwise it is the expected payoff conditional on the student's eligibility score. 
    \item The DA algorithm assigns students to colleges, where students' (ordinal) preferences are from Step 1, and colleges' lists of strict priorities are assembled in Step 2.
\end{enumerate}


\subsection{Strategic manipulation and incentives for disclosure}

We begin by assessing whether students have incentives to misreport rankings of colleges. A mechanism is \textbf{ordinally strategy-proof} if, for every student, for every auxiliary message behavior for her, and for all strategies of other students, it is optimal for her to report her ordinal preferences truthfully. This concept coincides with the standard strategy-proofness notion when a mechanism consists of only ordinal preferences as message space.

Our first main result outlines how the mechanisms of DA with voluntary information inherit a core property of strategic manipulability from DA:

\begin{proposition}\label{proposition:strategyproof}
All three mechanisms satisfy ordinal strategy-proofness.
\end{proposition}

In other words, Proposition \ref{proposition:strategyproof} states that partial truth-telling is a weakly dominating strategy in DAVID-Q and DAVID-U. Thus, no student can benefit from manipulating the order of colleges in her application. While this may sound weaker than the strategy-proofness of DA, it nevertheless elicits students' true ordinal ranking of colleges as in DA. 

We proceed by showing how DAVID-Q is safer than DAVID-U, because DAVID-Q prevents situations where students risk receiving lower utility from disclosing private information.
Suppose that a student $i$ submits a ROL that is a partial ordering of her true preferences, and let $\mu$ denote the matching produced by either the mechanisms DAVID-U and DAVID-Q. We say that the mechanism is \textit{disclosure safe} if the utility from matching, $v_{i,\mu(i)}$, cannot decrease if student $i$ discloses information to an additional college. 
We can now express our first result on the quota-based mechanism:

\begin{theorem}\label{theorem:DAVID-Q_safe}
Disclosure safety holds under DAVID-Q but not under DAVID-U.
\end{theorem}

\subsection{Justified envy}

Another core strategic property is whether participation in the matching mechanism prevents students obtaining a better match after assignment. We begin with outlining two variants of the justified envy property, one of which is new and incorporates the disclosure step by students. 

A mechanism eliminates \textbf{justified envy} if, for every preference and priority profile, there exists no pair of a student and a college such that the student prefers the college over her assignment and the college either has a vacant seat or the student's priority score exceeds that of an admitted student. The justified envy property extends naturally to a setting with multiple admission criteria where the property then has to hold at the sub-college level.
We amend the standard concept of justified envy by allowing for information disclosure through the submission of messages in addition to preferences. We say that a mechanism eliminates \textbf{justified envy with disclosure} if for every preference, auxiliary message and priority profile submitted to the mechanism there exists no pair of a student and a college such that the student prefers the college over her assignment and the college either has an empty seat or prioritizes the student higher than at least one admitted student. 
We note that the allowing for disclosure requires a stronger concept of stability.\footnote{The stronger stability criterion follows from the fact that justified envy with disclosure requires robustness against two kinds of blocking moves between students and colleges; one without and with the disclosure.}

Our next result outlines how the justified envy properties extend to mechanisms that allow for voluntary information. 

\begin{theorem}\label{theorem:justifiedenvy_davi}
The mechanisms satisfy the following justified envy properties. a) DA eliminates justified envy; b) DAVID-Q eliminate justified envy within a given admission quota, and; c) DAVID-U eliminates justified envy with disclosure.
\end{theorem}

We have completed our analysis of strategic properties. Both new mechanisms inherit strategy-proofness from DA. On the one hand DAVID-Q is disclosure safe. Yet, on the other hand,  DAVID-U fulfills the stronger criterion of justified envy and DAVID-Q only satisfies justified envy within quotas.

\subsection{Preference revelation equilibria}
We proceed by outlining the preference revelation games for each of our three mechanisms, which we use as our main equilibrium concepts. For tractability we focus mainly on large economies because under mild conditions on the distribution of agent types, it holds that there is a unique equilibrium.  We consider a continuum $E$ economy where the number of students has unit mass. We maintain that the set  $\mathcal{C}$ colleges is finite and thus, each college, $c$, has a mass of available seats denoted by $S_c$. In the remainder of this section, we solely describe students in terms of their type $\theta\in\Theta$.

\paragraph{Preference revelation games}

In each of the preference revelation games the students are the players and their utility is as outlined in the model. Moreover, we assume students know the exact distribution $\eta$ of other students' types, which makes it irrelevant to distinguish whether information is complete or incomplete. Below we outline the set of actions for agents in the three games. We denote the action of each student type by $\alpha_\theta\in A^m$, where $A^m$ is the set of actions for each student. Also, we define students' action profiles by $\alpha\in \mathcal{A}^m$, where $\mathcal{A}^m=(A^m)_{\theta\in\Theta}$. 

The action set for students in the preference revelation game of DAVID-Q, $A^{DAVID-Q}$,  is straightforward. A student's action is her rank-ordered list over sub-colleges and her set of actions is the feasible combinations of such lists. We also assume that students are indifferent about admission through the two quotas. As a consequence we can restrict rank-ordered lists over sub-colleges without a loss of generality as follows: regular sub-colleges are always included in the rank-ordered list whenever a student applies to information disclosure sub-colleges, and regular sub-colleges are ranked ahead of their disclosure sub-colleges.
We note that as DA is nested within DAVID-Q where we exclude the disclosure sub-colleges and set the disclosure quotas to zero, the action set $A^{DA}$ also consists of all feasible rank-ordered lists of colleges. 

In DAVID-U, the set of actions is different since a student's action consist of submitting both their rank-ordered lists and disclosed information to colleges: $A^{DAVID-U}=A^{DA}\times \{0,1\}^{C}$. 

\paragraph{Supply and demand framework}
We now extend the conceptual framework of supply and demand of \citet{Azevedo2016AMarkets} to our preference revelation games. For each action profile there is an associated assignment, $\mu$, and for DAVID-Q there is also a sub-college assignment, $\mu^{sub}$.

We begin with cutoffs that play the role of prices in markets. 
For a given matching $\mu$, we define \textbf{cutoffs} as minimal entry requirements  $P_\mu\in \mathcal{P}$ for admission at each college. For DA, the cutoff is the minimum eligibility score among those admitted. For DAVID-Q, the cutoffs at regular sub-colleges, $P^r$, are the minimum eligibility scores among those admitted, but for disclosure sub-colleges, the cutoff is the minimum payoff to the college among those admitted, $P^d$. We denote the profile of cutoffs at the sub-college level as $P$. For DAVID-U, there is a single, joint cutoff in terms of expected payoff to the college, which dictates a uniform entry requirement regardless of whether information disclosure is used or not. When the measure of students admitted at a college is lower than the measure of seats there, then any student can gain admission there, which we denote by $P_c=-\infty$. We let $\mathcal{P}^m$ denote the space of feasible cutoffs at mechanism $m$, which for DA and DAVID-U equals $\mathbb{R}^C$ while for DAVID-Q it equals $\mathbb{R}^{2C}$.

We say that student $\theta$ can \textbf{afford} college $c$ under cutoffs if the college is willing to admit $\theta$. The willingness to admit under DA requires that $e_{i,c}>P_{c}$ and under DAVID-Q it requires that $e_{i,c}>P_{c}^{r}$ for regular admission or $w_{i,c^d}>P_{c}^{d}$, while for DAVID-U it requires that $E[w_{i,c}|e_{i,c}]>\Pi_i$ for regular admission or $w_{i,c}>\Pi_c$ for admission with disclosure.

A student’s \textbf{demand} given cutoffs is her favorite college in terms of utility after subtracting the associated applications costs for admission among those she can afford.\footnote{There are zero application costs for regular admission but admission with disclosure in DAVID-Q and DAVID-U is associated with a cost of $\gamma_{\mathcal{I}}$.} The mathematical definition of demand functions is found in Appendix~\ref{app:demand_exact_definition}. An implication of our definition of demand is that if students of type $\theta$ demand a college $c$ given cutoffs $P$ then they expect admission there. If no colleges are affordable, define $D^m_\theta(P)=\emptyset$, meaning that the students' demands are unmatched. We say that demand occurs with  \textbf{disclosure} if admission requires disclosure. Finally, we define the  \textbf{aggregate demand} for each college, $c$, over students as:

\begin{align}
    D_c(P)=\int\eta(\theta)\,\textbf{1}[D^\theta(P)=c]\,\text{d}\theta
\end{align}

Also denote the $D^m$ as the vector of aggregate demands for the colleges in $\mathcal{C}$ under mechanism $m$. In addition, let $D_{\theta,c}^m=1$ if $D_{\theta,c}^m=c$, otherwise $D_{\theta,c}^m=0$.

In the remainder of this subsection we denote the dependence of the demand on measures as $D(\cdot|\eta)$. An essential condition for uniqueness of our measure is the smoothness of the demand measures:

\begin{definition} The distribution of student types $\eta$ is regular in mechanism $m$ if the image under $D^m(\cdot|\eta)$ of the closure of the set \fixme{not defined for $\Pi$ yet} \[\{P\in\mathbb{P}^m: \,D^m(\cdot|\eta) \mbox{ is continuously differentiable at }P\}\] has Lebesgue measure 0.\end{definition}

Examples of sufficient conditions for regularity of $\eta$ includes $D(\cdot|\eta)$ being continuously differentiable or if $\eta$ has a continuous density.

\paragraph{Equilibrium concepts and properties}
The equilibrium associated with each of the three preference revelation games above requires that students play their best respond to other agents' actions. 
We note that the individual demand does in fact capture the best response in terms of submitted preferences and targeting behavior for every agent given cutoffs, see Appendix~\ref{app:demand_exact_definition} for the definition of demand. From these definitions of demand we note that individual demand is, by construction, a pure strategy as preferences are strict. 

We now analyze the equilibria in large economies, which allows us to pin down uniqueness properties under weak sufficient conditions.\footnote{We note that Theorem~\ref{theorem:unique_eqilibria} is almost identical to \citet{Azevedo2016AMarkets}. Thus, our proof demonstrates how their results can be generalized and extended to our setting.}

\begin{theorem}\label{theorem:unique_eqilibria}
Consider an economy $E = [\eta, S]$.\begin{enumerate}
    \item If $\eta$ has full support for mechanism $m$, then the economy $E$ has a unique equilibrium in the preference revelation associated with $m$.
    \item If $\eta$ is any regular distribution for mechanism $m$, then for almost every vector of capacities, $S$, with $\sum_{c\in \mathcal{C}}S_c < 1$, the economy, $E$, has a unique equilibrium for the preference revelation game associated with $m$.\end{enumerate}
Moreover, if there is a unique equilibrium under DAVID-Q and DAVID-U, then there exists a disclosure share such that the equilibria coincide.
\end{theorem}

The above result shows that we can pin down the unique equilibrium and its associated matching exactly under one of two conditions. First, if there is full support such that every agent type has positive measure but still zero mass. Second, even if full support is violated, then if the measure of type is regular then uniqueness still holds. 

We finalize our analysis of equilibria by noting that, in finite economies, there may not be an equilibrium in the preference revelation game for DAVID-U due to the way we have defined demand.

\begin{example}\label{example:no_combined_stability}
Let there be two students denoted by 1,2 and one college with a single seat, which we omit notation for. We assume that $v_1,v_2>\gamma_\mathcal{I}$ as well as $e_1>e_2>0$, $w_1>w_2>0$ and $w_2>e_1$. Both students will report they would like to match with the college. Suppose student 1 targets the college and that student 2 does not target the college. Student 1 can stop targeting the college to increase her utility, which will lower the admission cutoff and, in turn, student 2 can target the college and thus destabilize the equilibrium. After losing the seat, student 1 may gain from targeting the college, which will lead student 2 to not target it. We have now arrived at where we started, which implies that there is no equilibrium.
\end{example}

Although potential non-existence of an equilibrium is an issue for the matching environment, we note that that the issue may be fixed by using a stability concept with additional refinement. One way is to let agents be farsighted, which means that students will keep targeting colleges where they can gain admission without targeting. The reason is that these agents realize that stopping targeting the college leads to cycles of instability. As a consequence, agents will not block the matching outcome as there exists a path leading back to the original outcome in which the deviators are not better off \citep{Mauleon2011VonMatching}.

\section{Welfare analysis of voluntary information disclosure}

We proceed by analyzing the welfare implications of allowing for voluntary information disclosure. We focus on large economies as we demonstrated unique equilibria and associated matching outcomes. We perform our analysis by comparing the welfare for students and colleges separately under DA and DAVID-U.

To simulate welfare of DA and DAVID-U we specify the joint probability distributions for student types. To this end, we assume student types are generated in a two stage process. Initially, we draw a student type 'anchor' for each type denoted by $\theta_i=(e_i,w_i,v_i)$. We let $\theta_i$ follow a multivariate Gaussian distribution. This allows us to compare welfare under variations of correlation between $e_i$, $w_i$, and $v_i$. One very important feature is the fact that a student type anchor for student $i$ is identical across every college for student $i$. Therefore, $\theta_i$ only varies across students. In the second step we want to allow for individual idiosyncratic taste and payoff shocks across colleges. We therefore introduce an idiosyncratic error for $v_{i,c}$ and $w_{i,c}$ and let them follow a multivariate Gaussian distribution. To mimic real-life college choice mechanisms with a single eligibility score across colleges, we set the idiosyncratic error for eligibility score to zero. We add the idiosyncratic taste shocks to $w_i$, and $v_i$ to achieve the student-college specific values of $w_{i,c}$ and $v_{i,c}$. 

Recall that colleges prioritize students according to expected college payoff conditional on the information available. If a student does not disclose information, the following formula is used to calculate the expected college payoff:\footnote{$\mu_w$ and $\mu_e$ are the expected value of respectively $w_i$ and $e_i$. $\rho_{w,e}$ is the correlation between $w_i$ and $e_i$. $\sigma_w$ and $\sigma_e$ are the standard deviations of $w_i$ and $e_i$, respectively.}
\begin{align*}
    E[w_{i,c}|e_{i,c}=e]=\mu_w+\rho_{w,e}\frac{\sigma_w}{\sigma_e}(e-\mu_e)
\end{align*}

The final requirement to conduct simulations of welfare is to specify student behavior. For DA matching we assume students report truthfully, since this is a weakly dominating strategy. For the DAVID-U matching, we compute an equilibrium by iteratively updating their demand, which is introduced in a previous section:

\begin{enumerate}
    \item Students submit a complete and truthful ranking of colleges without disclosure (DA matching). This produces initial cutoffs.
    \item Students update their demand according to the initial cutoffs produced by the initial matching above, possibly by targeting colleges.
    \item The previous step is repeated  based on the previous cutoff until the college cutoff levels do not change anymore and a market clearing cutoff has been reached. The final match is saved as the DAVID-U matching.
\end{enumerate}

\subsection{Student Welfare}

\begin{figure}[!ht]
    \begin{subfigure}[b]{.495\linewidth}
        \includegraphics[width=\linewidth]{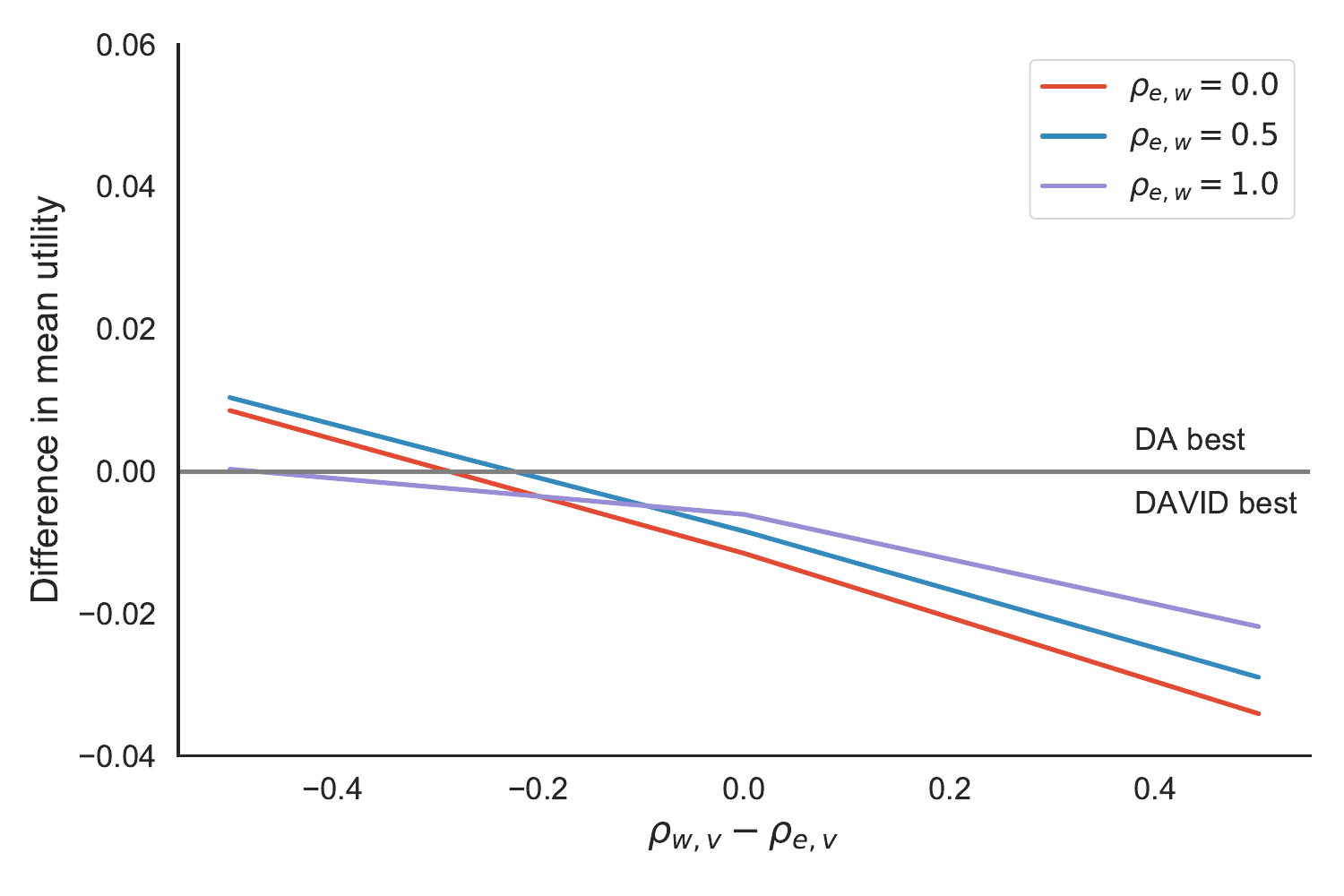}
        \caption{No application costs, ($\gamma_I=0$)}
        \label{fig:mc_stud_welfare_A}
    \end{subfigure}
    \begin{subfigure}[b]{.495\linewidth}
        \includegraphics[width=\linewidth]{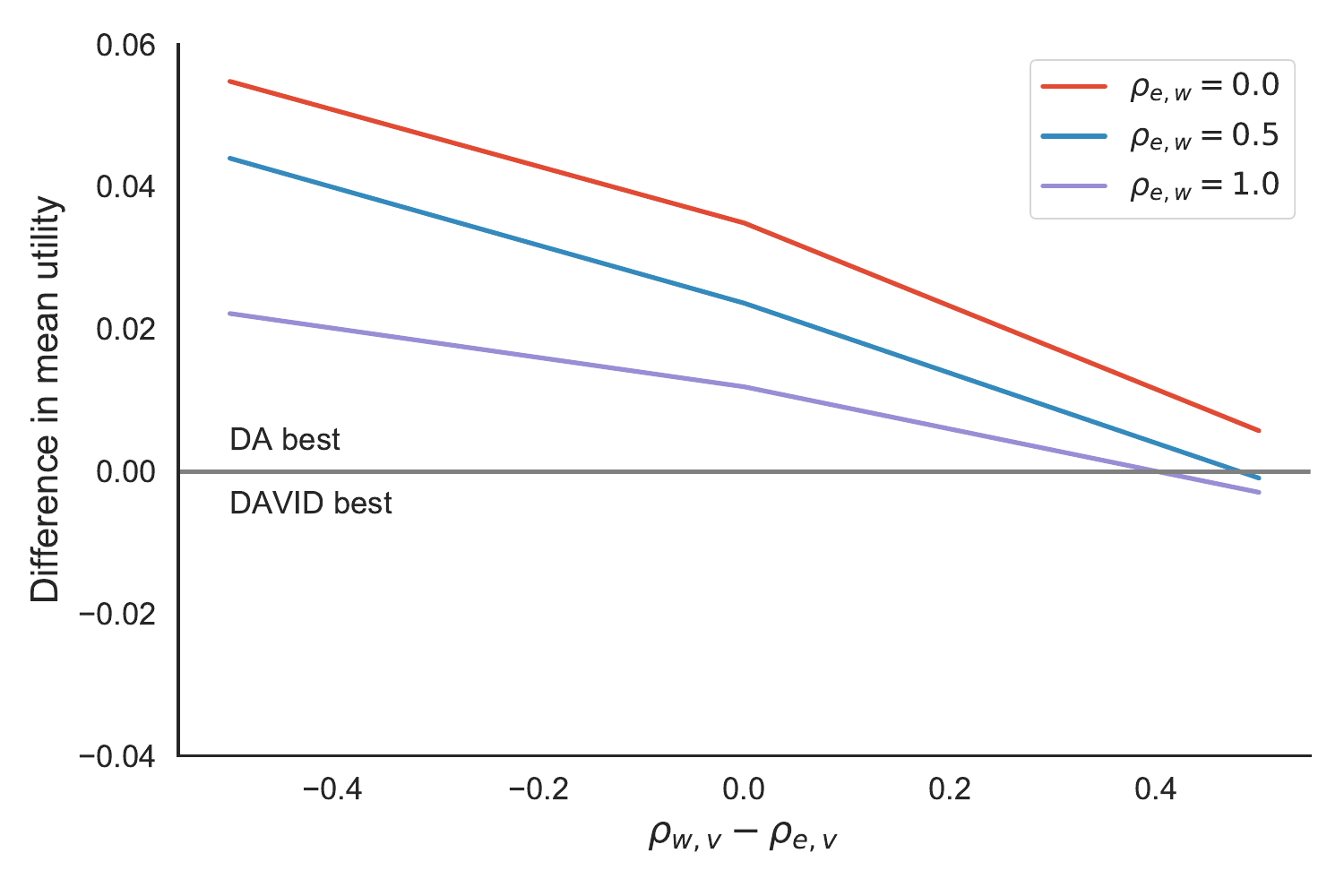}
        \caption{With application costs, ($\gamma_I=0.05$)}
        \label{fig:mc_stud_welfare_B}
    \end{subfigure}
\caption{Comparison of student utility under DA and DAVID-U}
\label{fig:mc_stud_welfare}
\floatfoot{Notes: The two panels above contain figures illustrating the difference in average net student utility under DA and DAVID-U matchings with various parameter values. If a curve is below (above) 0, DAVID-U has a higher (lower) average net student utility than DA. The horizontal axis shows the difference between the correlation of student utility and college payoff and the correlation of eligibility score and student utility. The left side figure has no application cost. The right side figure has a positive application cost of 0.05. The parameter values not shown in the figures are as follows: 500 simulations, 100 students, 6 colleges, 10 seats at every college, $\gamma_C=0$, $\rho_{w,v}=0.5$, $\mu_e=1$, $\mu_w=1$, $\mu_v=1$, $\sigma_e=0.1$, $\sigma_w=0.1$, $\sigma_v=0.1$, and std of $\epsilon_w$ is 0.1.}
\end{figure}

We assess student welfare by calculating the average net student utility\footnote{Net student utility accounts for application cost} for DA and DAVID-U matching. Our primary focus is the qualitative difference in student welfare under DA and DAVID-U. As a result, we present the difference in average net student utility under DA and DAVID-U in Figure~\ref{fig:mc_stud_welfare}.  The figure includes several plots to cover welfare implications under key parameter values. We briefly state the key parameters and a motivation for including them: \\

\noindent \textbf{Application cost ($\gamma_I$):} This parameter affects students' utility through the costs of targeting colleges. We let the costs vary between 0.00 and 0.05. The latter represents 5 percent of the mean utility from entering a college. \\

\noindent \textbf{Correlation between eligibility score and college payoff ($\rho_{e,w}$):} This parameter captures the informativeness of eligibility scores towards college payoff and thus does not directly determine students' utility. However, it affects the degree to which admission under DAVID-U is different from DA due to changes in colleges' priorities and thus has indirect implications for student welfare. \\

\noindent \textbf{Difference in correlation with utility for college payoff and eligibility score ($\rho_{w,v}-\rho_{e,v}$):} The correlation between eligibility score and student utility is a measure of the correlation between public information about students and students' preferences. Similarly, the correlation between college payoff and students' preferences is a measure of the correlation between private information about students and students' preferences. Hence, this metric illustrates how a relatively stronger correlation between private information and students' preferences affects the relative student welfare gain under DAVID-U. \\


We show student welfare under different combinations of parameters above in Figure~\ref{fig:mc_stud_welfare}. For each subplot, the vertical axis measures how attractive DA is relative to DAVID-U. Similarly, for each subplot a move along the horizontal axis implies that correlation of college welfare and student utility increases compared to the correlation of eligibility score and student utility. 

We draw three conclusions on how parameters affect student welfare by inspecting Figure~\ref{fig:mc_stud_welfare}.
First, higher application costs makes DAVID-U less favorable. This is seen by the observation the curves are above 0 in the right hand figure where students incur application costs. Second, higher correlation between student utility and college payoff makes DAVID-U more favorable when the correlation of student utility and eligibility score is kept fixed. This is seen from the negative slopes in all panels.
Finally, DAVID-U is more favorable to students when correlation between eligibility score and college payoff is low, as this means eligibility score has little informational value about colleges' payoffs. This is seen from comparing the different curves in both figures, where the upper curves have higher student welfare.

\subsection{College Welfare}
\begin{figure}[!t]
    \begin{subfigure}[b]{.495\linewidth}
        \includegraphics[width=\linewidth]{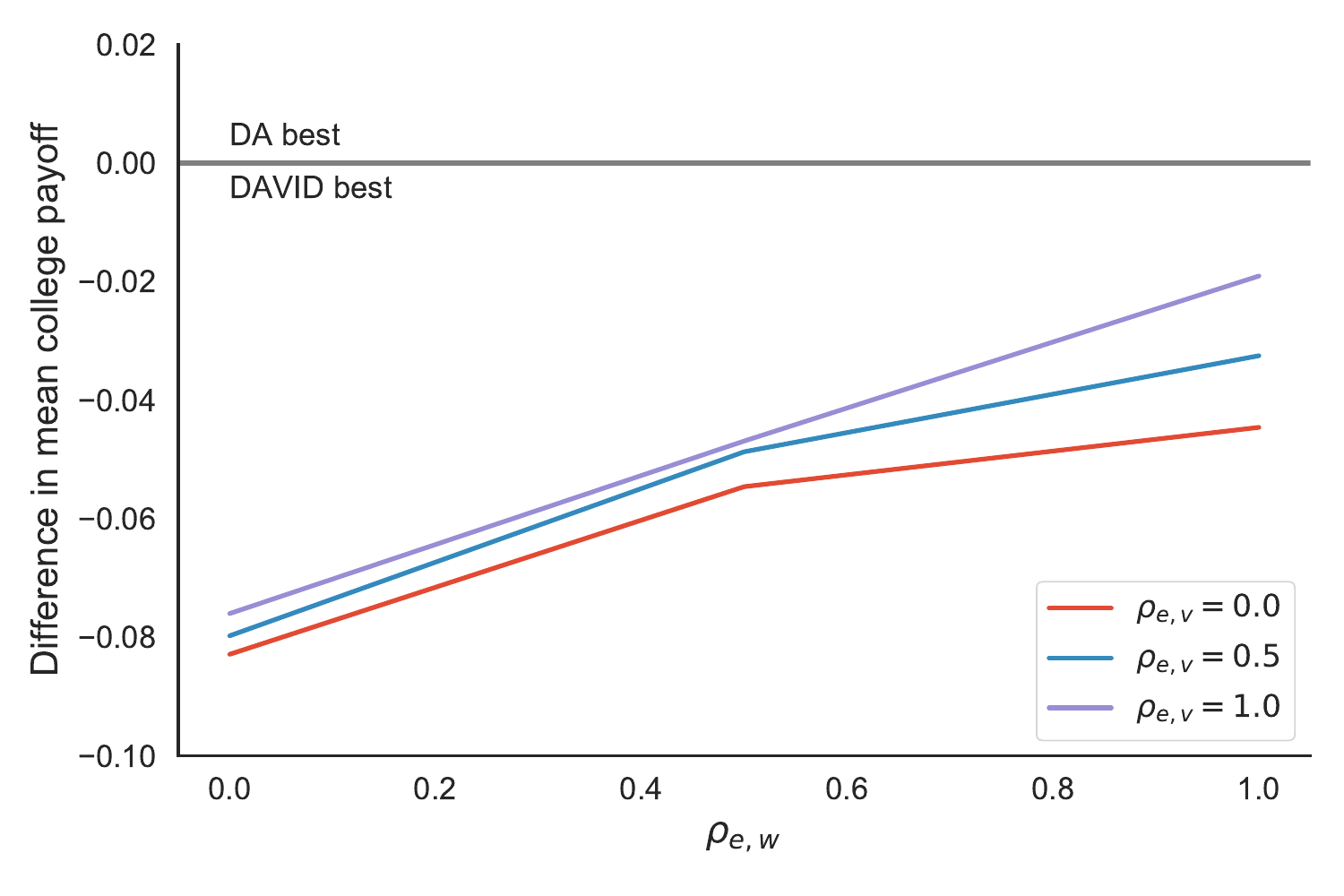}
        \caption{No screening cost ($\gamma_C=0$)}
        \label{fig:mc_college_welfare_A}
    \end{subfigure}
    \begin{subfigure}[b]{.495\linewidth}
        \includegraphics[width=\linewidth]{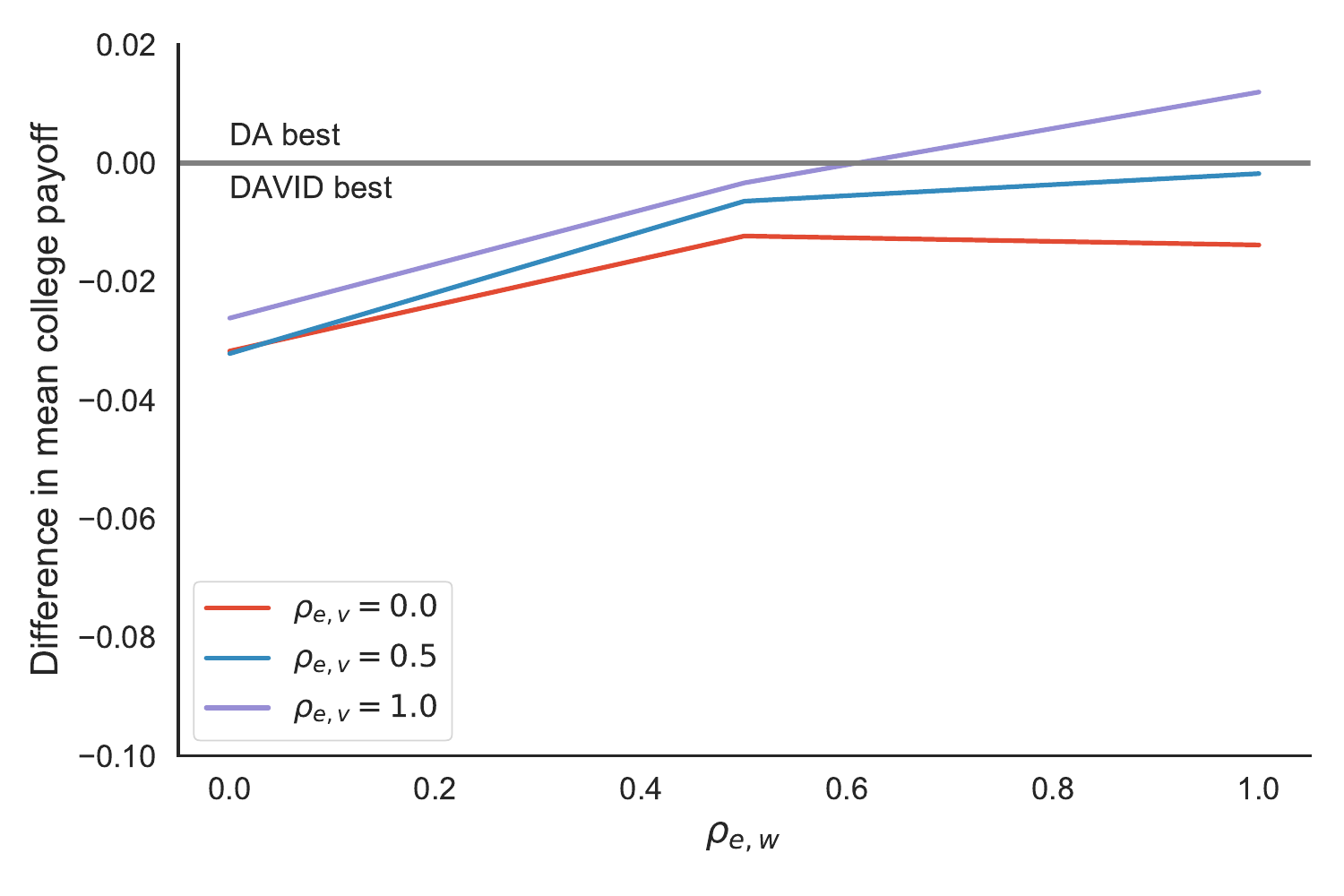}
        \caption{With screening cost ($\gamma_C=0.05$)}
        \label{fig:mc_college_welfare_B}
    \end{subfigure}
\caption{Comparison of college payoff under DA and DAVID-U}
\label{fig:mc_college_welfare}
\floatfoot{Notes: The two panels above contain figures illustrating the difference in average net college payoff under DA and DAVID-U matching. If a curve is below (above) zero, DAVID-U has a higher (lower) average net college payoff than DA. The horizontal axis shows the correlation of eligibility score and college payoff. The figure on the left side has no screening costs. The figure on the right side has positive screening costs. The parameter values not shown in the figures are as follows: 500 simulations, 100 students, 6 colleges, 10 seats at every college, $\gamma_I=0.05$, $\mu_e=1$, $\mu_w=1$, $\mu_v=1$, $\sigma_e=0.1$, $\sigma_w=0.1$, $\sigma_v=0.1$, and std of $\epsilon_w$ is 0.1.}
\end{figure}

We now assess how colleges' welfare is affected by using voluntary screening. We measure college welfare as the average net college payoff from an accepted student\footnote{Net payoff for a college accounts for its cost of screening information for every student that has disclosed information to it.} under matching with DA and DAVID-U. Our primary focus is the qualitative difference in student welfare under DA and DAVID-U. Figure \ref{fig:mc_college_welfare} illustrates the difference in average net college payoffs under DA and DAVID-U. The figure includes several plots to cover welfare implications under key parameter values. We briefly state the key parameters and a motivation for including them:\\

\noindent \textbf{Screening cost ($\gamma_c$):} We vary screening costs between none and 0.05. The latter represent 5 percent of the mean college payoff from accepting a student. Screening costs are included because they decrease the net college payoff in DAVID-U.\\ 

\noindent \textbf{The correlation between eligibility score and student welfare ($\rho_{e,w}$):} Same motivation as for the student welfare analysis.\\

\noindent \textbf{The correlation between eligibility score and student utility ($\rho_{e,v}$):} This metric is included because it has a high influence on the behavior of the students. If the correlation is very high, students are less likely to disclose information to colleges at which they have a higher college payoff.\\

Figure \ref{fig:mc_college_welfare} shows subplots with variations of the parameters described above. Similarly, for each subplot a move along the horizontal axis implies that the correlation of eligibility score and college payoff increases. In both figures a move along the horizontal axis decreases colleges payoff by using DAVID-U instead of DA. We draw two conclusions on how parameters affect college payoffs by inspecting the figure. First, DAVID-U is more favorable, when the correlation between eligibility score and college payoff is low. This is seen from the curves in Figure ~\ref{fig:mc_college_welfare}, which have a positive slope. This makes intuitive sense as lower correlation means that the eligibility score is a poor indicator of student worth and thus increases the value of screening. Second, DAVID-U is more favorable than DA for colleges without screening costs and in certain cases with screening costs. The additional requirement for DAVID-U to be favorable is that correlation between eligibility score and payoff is no more than 0.5, see Figure~\ref{fig:mc_college_welfare_B}.

\section{Empirical evidence from Denmark}

We round off our analysis by examining behavioral patterns of students who participate in DAVID-Q in an empirical context by leveraging Danish admission data. We aim to empirically test core theoretical properties and behavioral patterns predicted by DAVID-Q and provide evidence to support voluntary information disclosure in matching markets. To be specific, we set up \fixme{three} two testable hypotheses. However, before we introduce the \fixme{three} two hypotheses, we briefly explain the system for admission to higher education in Denmark and how it relates to DAVID-Q.

\paragraph{Admission to higher education in Denmark} 

The annual enrollment to Danish undergraduate programs consists of approximately 90,000 students and 900 study programs. Each student is allowed to apply to a maximum of 8 study programs. Each study program is allowed to admit students through two separate quotas, quota 1 and quota 2, resembling a DAVID-Q setup. The study programs submit their quota-specific capacities to a central clearance house before the programs receive any disclosed information. In quota 1, students are prioritized according to an eligibility score. The eligibility score is a weighted average of a student's high school grades and it is identical across every study program. In quota 2, students are prioritized according to disclosed student information and discretionary study program rules. The study programs vary in how much information they require students to disclose. Examples are motivational letters, in-person interviews, and multiple choice tests. The timeline for admission is as follows. The deadline for quota 2 applications is March 15. The deadline for quota 1 applications is July 5. The combined matching of quota 1 and 2 applications takes place during the period of July 5 to July 27. The students receive admission offers on July 28. There is an aftermarket that offers first-come-first-served available seats at study programs in August. 
After the main assignment has taken place, The Ministry of Higher Education publishes eligibility cutoff levels for quota 1 admission to every study program on July 28. This information receives a great deal of public attention and it remains available online in forthcoming years. As a result students can access historical study program cutoff levels as far back as 1977.

\begin{figure}[!ht]
    \begin{subfigure}[b]{.4\linewidth}
        \includegraphics[width=\linewidth]{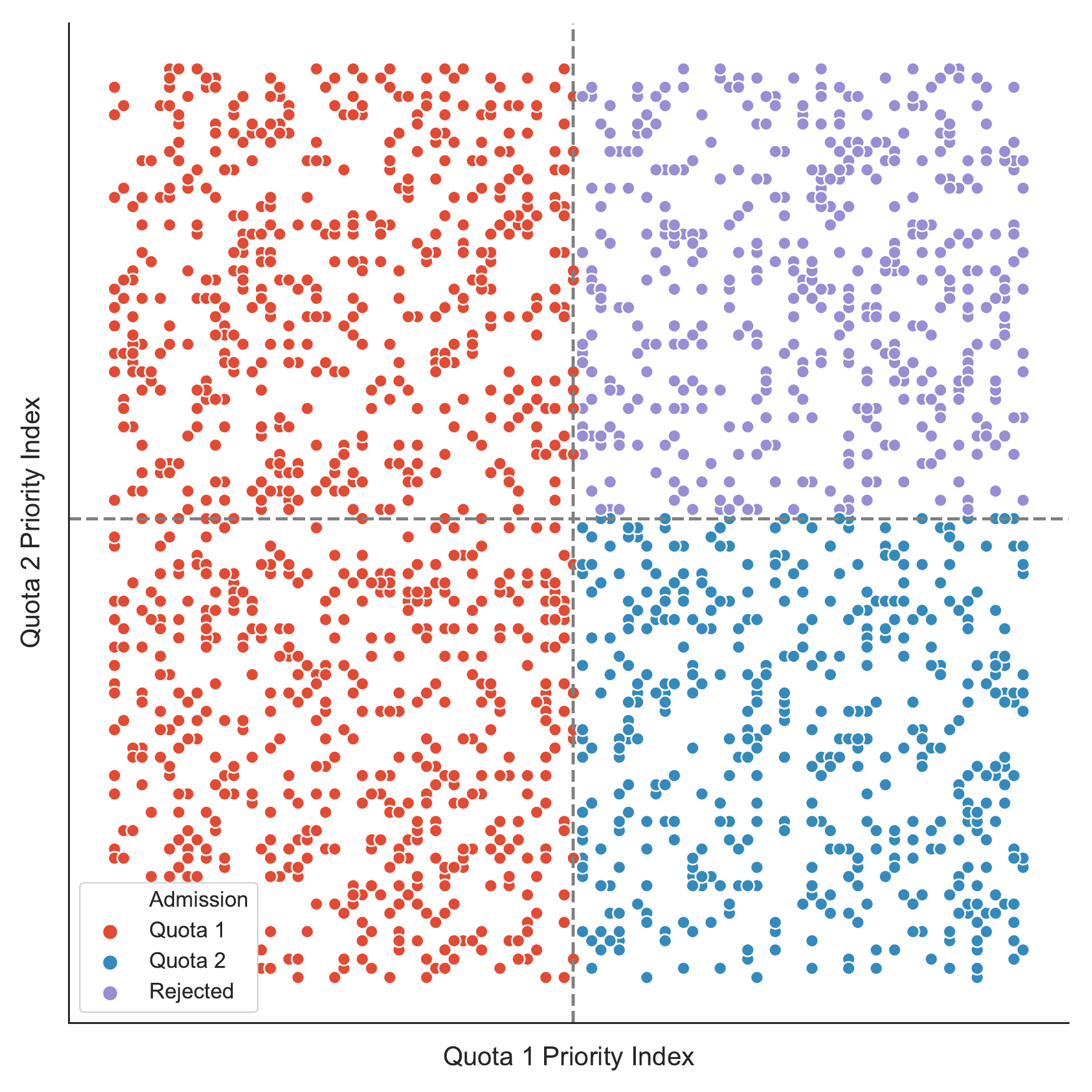}
    \end{subfigure}
\caption{Dual admission quotas: quota 1 (eligibility) and quota 2 (information disclosure)}
\label{fig:illustration_of_quota_adm}
\floatfoot{Notes: This figure illustrates how a study program admits student through quota 1 and quota 2 in the system for admission to higher education in Denmark. The horizontal axis contains students' quota 1 priority index at a study program. That is, how a study program prioritize a student. The quota 1 priority index is a direct mapping of a student's eligibility score. That is, the student with the highest eligibility score among potential students to a program receives a priority index of zero. Similarly, the vertical axis contains students' quota 2 priority index at a study program. E.g., if a student has a low (high) quota 2 priority index it implies that a program has a high (low) priority of that student. The dotted lines illustrates the priority index required to gain admission (cutoff levels). If a student has a lower quota 1 priority index than the quota 1 cutoff level, she is admitted through quota 1. These students are indicated by red dots. Similarly, if a student is not admitted through quota 1 and has a lower quota 2 priority index than the quota 2 cutoff level, she is admitted through quota 2. These students are indicated by blue dots. Finally, students whose quota 1 and 2 priority indices are above the cutoffs are rejected. These students are indicated by purple dots.}
\end{figure}

To explain our hypotheses, we illustrate quota admission in Figure~\ref{fig:illustration_of_quota_adm}. In this figure a dot represents a student who has applied to a specific program through quota 1 and 2. That is, the dots represent students who voluntarily disclose information. The red dots represent students who have a low enough priority index to be admitted to the program through quota 1. The blue dots represents students who has a low enough priority index to be admitted through quota 2. Notice, that these students colored in blue would not have been admitted if voluntary information disclosure was not an option.   

In our theoretical model, we assume colleges are capable of using disclosed information to prioritize students according to their underlying payoff to colleges. However, in a real world setting disclosed information may still contain noise that prevents colleges from deriving underlying payoffs. Therefore, our first hypothesis seeks to measure whether disclosed information actually improves how colleges prioritize student applications.

\begin{hypothesis}
\label{hypothesis:program_use_disclosed_information}
Colleges use disclosed information to prioritize students who provide them with an increased payoff. Since colleges are paid for student completion, we can operationalize this as higher ranked students have higher graduation rates.
\end{hypothesis}
Our second hypothesis tests the assumption that voluntary information disclosure is used by the students it was designed for. More specifically, the students who need a second chance of admission because their eligibility score is too low for admission through a regular admission quota. 
This is an important welfare aspect because programs' screening costs would be very high if every student disclosed information. The hypothesis is mathematically also predicted by our demand function for DAVID-Q that we outline in Appendix~\ref{app:demand_exact_definition}.

\begin{hypothesis} \label{hypothesis:rational_targeting}
Among applicants to a degree program, a higher eligibility score leads to a lower propensity of disclosing information. 
We operationalize this as a decreasing propensity to disclose information in students' highest ranked study program for increased differences between her own eligibility score for the program and the public available admission cutoff from the previous year. 
\end{hypothesis}

\paragraph{Analysis}

We begin by examining empirical evidence for Hypothesis~\ref{hypothesis:program_use_disclosed_information}. We approach this by comparing the graduation rate of information disclosing students across their quota 2 priority index. That is, whether students who are highly prioritized by programs are more likely to graduate than students with a lower priority. We apply a regression approach and estimate the following linear probability model:   

\begin{align} 
y_i = \sum_{q=2}^{Q=4}\alpha_q Q2PriorityIndexQuartile +  \gamma_i + \epsilon_i  \label{eq:hypothesis_1}
\end{align} 
where the outcome variable $y_i$ is a student's graduation rate. Our parameter of interest is $\alpha_q$ which measures the difference in the graduation rate between students in the first quartile of quota 2 to students in lower quota 2 quartiles, $Q2PriorityIndexQuartile$. We include a wide range of fixed effects. Specifically, $\gamma_i$ contains fixed effects for year of admission, study program, eligibility score, and interaction between study program and eligibility score. We do not include a student's rank of a program, gender or age, because this information is not available to study programs prior to admission. We show the estimated coefficient results in Figure~\ref{fig:HYPOTHESIS_1}, see Table~\ref{tab:graduation_rate_priority} in Appendix~\ref{app:tables} for detailed regression estimates. We estimate that the graduation rate is approximately 8 percentage points higher for students in the first quartile of quota 2 priority indices compared to students in the fourth quartile. This corresponds to a 13 percent increase in the graduation rate when comparing the first quartile with the fourth quartile. 

Next, we examine the empirical evidence for Hypothesis \ref{hypothesis:rational_targeting}. 
We do so by considering the share of students who disclose private information. Under the assumption that students use last years' admission cutoff to form admission beliefs, we can use the distance between a student's eligibility score and last year's cutoff of their top-ranked study program as a proxy for admission beliefs. E.g. if a student's eligibility score is far below (above) last year cutoff of her top-ranked program, we assume she believes her admission prospect without targeting is very low (high). Figure \ref{fig:disclose_info_score} depicts the share of targeting students conditional on the distance between eligibility score and last year's program cutoff. From this figure we can tell that the propensity to target a top-ranked program indeed declines when a student is more likely to be admitted without targeting. Hence, we conclude from our inspection of Figure~\ref{fig:disclose_info_score} that there is empirical evidence in favor of Hypothesis \ref{hypothesis:rational_targeting}.

\begin{figure}[!ht]
    \begin{subfigure}[b]{.47\linewidth}
        \includegraphics[width=\linewidth]{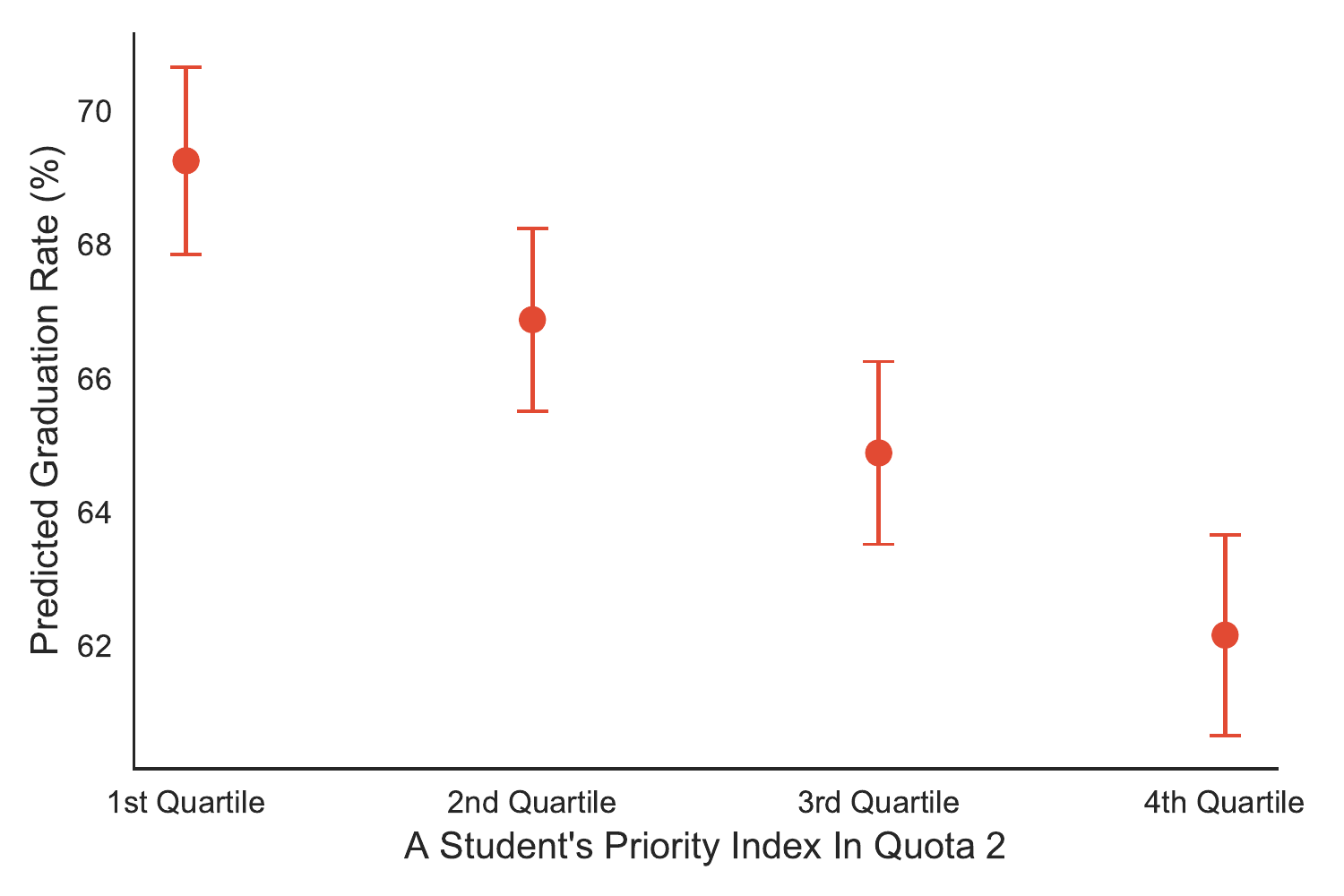}
        \caption{College ranking of students by disclosure and graduation rate}
        \label{fig:HYPOTHESIS_1}
    \end{subfigure}
    \hfill
    \begin{subfigure}[b]{.47\linewidth}
        \includegraphics[width=\linewidth]{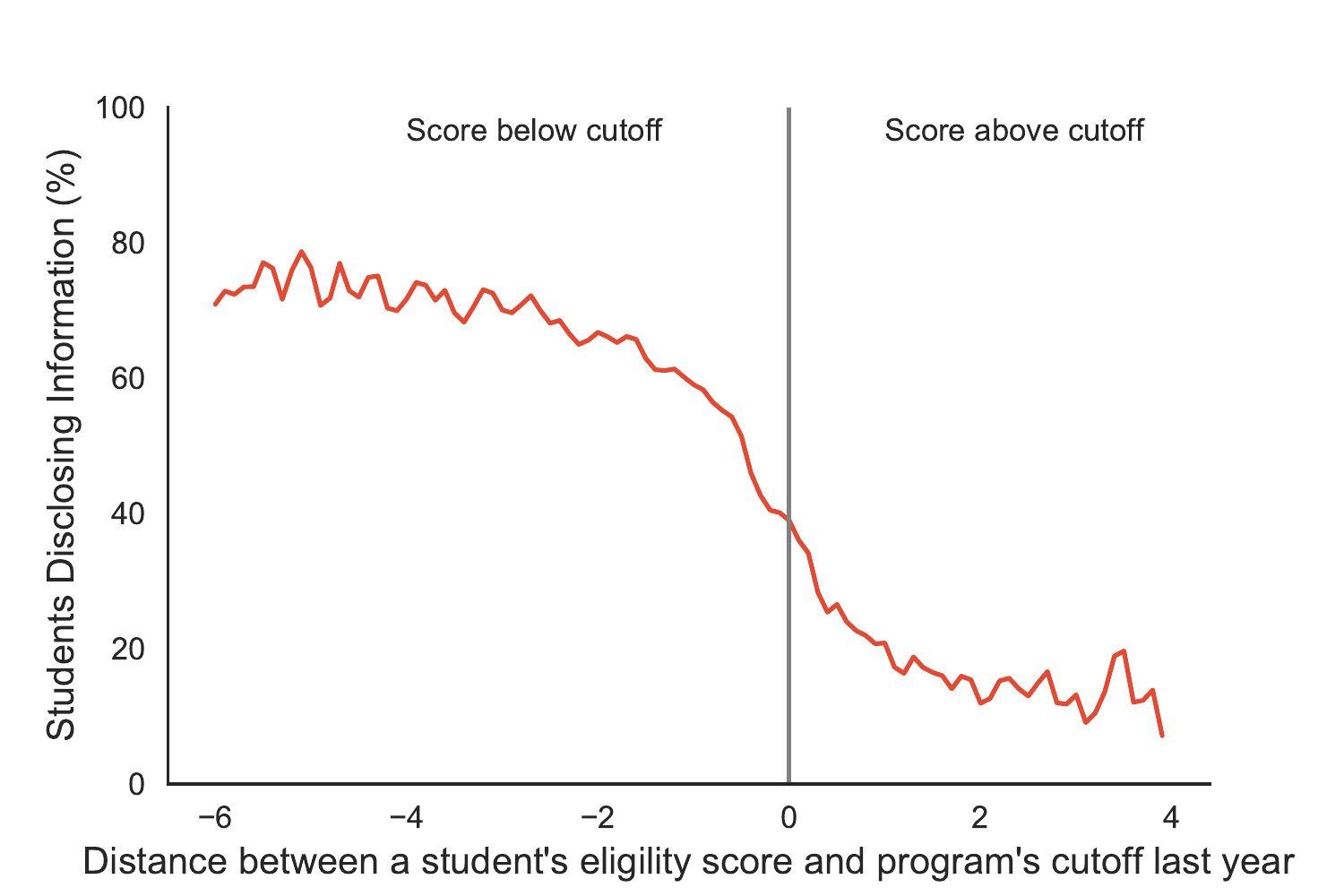}
        \caption{Students' Propensity to Disclose Information and Eligibility Score cutoff gap}
        \label{fig:disclose_info_score}
    \end{subfigure}
\caption{Examination of Hypothesis \ref{hypothesis:program_use_disclosed_information} and \ref{hypothesis:rational_targeting}}
\label{fig:empirical_evidence}
\floatfoot{Notes: The left-hand panel shows the regression results of a linear regression model. The right-hand panel shows the share of students who voluntarily disclose information conditional on the distance between a student's eligibility score and last year cutoff for the student's top-ranked program. The figure only considers students who apply for a study program with a cutoff above 8 for the previous year. The line is interpolated by using observations within 0.1 distance in terms of cutoff distance to compute the share of students who disclose information.}
\end{figure}

\fixme{incorporate selection hypothesis (needs motivation): To examine Hypothesis \ref{hypothesis:virtous_selection} empirically, we compare the graduation rate among admitted students by splitting on whether or not students disclosed information to the study program they were admitted to. The intuition being, that the graduation rate should be higher among disclosing students if disclosed information plays a role in unobserved matching effect. We define graduation as a student who either finished her program or remained enrolled four years after enrollment. We define the difference in graduation rate between the two groups as a disclosure effect. We estimate the following linear probability model}
\fixme{Her vil det være godt at nævne at problemet er et simpelt prediction policy problem - hvem skal man vælge - referer til mullainathans arbejde}

\section{Discussion}

There are a number of aspects of the application process that we have not considered in this paper. For instance, we assume that students have complete information about other students' types. Although such information is unlikely in practice, we conjecture that our results extend to a setting with incomplete information and application costs to submit preferences as in \citet{Fack2019BeyondAdmissions}.

Another assumption is that screening leads to the exact revelation of students' worth to colleges. In our empirical analysis we saw that screening only provides better predictions of their admission success, not exact knowledge of these outcomes. Thus, our conclusions would likely hold at the interim stage before matching. This would mean that application costs must be lower relative to possible matching quality gains for both students and colleges to benefit. This is likely a valid assumption in reality where one would expect that application and screening effort is outweighed by the possibility of better admission prospects. We also note that our simulation does not examine distributional consequences - it may be that allowing for voluntary disclosure increases assortative matching by ability into colleges, which may have adverse effects if there are peer effects.

Future work may want to generalize our framework for admission under multiple criteria where students self-select into which criteria they want to pursue admission under. To some extent, this notion generalizes earlier work on controlled school choice where quotas for certain distributional goals are used. Another route for future work is to extend our framework to quantify how uncertainty about admission prospects \citep[as in e.g.,][]{Fack2019BeyondAdmissions,} or about own preferences \citep[as in e.g.,][]{grenet2022preference} may affect welfare in our framework. Finally, it would be interesting to quantify and evaluate the welfare impact and strategic properties of voluntary screening either in experiments or from field studies.

\printbibliography[]

\newpage
\appendix

\section{Supply and demand in preference revelation games}

\subsection{Demand functions}\label{app:demand_exact_definition}

Here, we exhibit the individual demand by type $\theta$ function for a given cutoff under the three preference revelation games: 
\begin{itemize}
    \item For DAVID-Q,  $D_{\theta,c}^{DAVID-Q}(P^{sub})=\varepsilon_{\theta,c}^{DAVID-Q}(P^{sub})+\sigma_{\theta,c}^{DAVID-Q}(P^{sub})$, see equations \eqref{eq:admission_david_q_eligibility}-\eqref{eq:admission_david_q_disclosure_outside}.
    \item For DAVID-U,  $D_{\theta,c}^{DAVID-U}(\Pi)=\varepsilon_{\theta,c}^{DAVID-U}(\Pi)+\sigma_{\theta,c}^{DAVID-U}(\Pi)$, see equations \eqref{eq:admission_david_u_eligibility}-\eqref{eq:admission_david_u_disclosure_outside}.
\item For DA, $D_{\theta,c}^{DA}(P)=\varepsilon_{\theta,c}^{DA}(P)$, where only equations \eqref{eq:admission_david_q_eligibility} and \eqref{eq:admission_david_q_eligibility_outside} are satisfied and we let $P_c^d=\infty$ for all $c\in\mathcal{C}$.\footnote{This follows as DAVID-Q is equal to DA if we set the disclosure share to zero for all colleges.}
\end{itemize}

In what follows we describe $\varepsilon_{\theta,c}^{m}$ and $\sigma_{\theta,c}^{m}$, which are, respectively, with and without disclosure.  

We let $\varepsilon_{\theta,c}^{m}=1$ denote that students of type $\theta$ demand college $c$ without disclosure, otherwise $\varepsilon_{\theta,c}=0$ if they do not. Such demand  requires first that the student must be eligible for $c$ and must prefer admission there ahead of the null match, which is captured by \eqref{eq:admission_david_q_eligibility} and \eqref{eq:admission_david_u_eligibility}.  In addition, the student must have no better options at other colleges, which requires that $\hat{\varepsilon}_{\theta,c,c'}=1$ and holds if either admission is not feasible or not profitable if feasible, see \eqref{eq:admission_david_q_eligibility_outside} and \eqref{eq:admission_david_u_eligibility_outside}. 

Likewise we let $\sigma_{\theta,c}=1$ denote that student $i$ demands college $c$ with disclosure, which requires that the following analogous conditions hold. First, the student must be ineligible for $c$, but able to gain admission through disclosure, and must prefer admission at $c$ net of application costs ahead of the null match, which is captured in the first part of \eqref{eq:admission_david_q_disclosure} and \eqref{eq:admission_david_u_disclosure}. Moreover, the student must have no better options at other colleges, which requires that $\hat{\sigma}_{\theta,c,c'}=1$ and holds if either admission is not feasible or not profitable if feasible, see \eqref{eq:admission_david_q_disclosure_outside} and \eqref{eq:admission_david_u_disclosure_outside}. 
\begin{align}
\varepsilon_{\theta,c}^{DAVID-Q}(P^{sub})&=\textbf{1}\big(v_{\theta,c}>0\big) \cdot \textbf{1}\big(e_{\theta,c}>P^{r}_c\big)\cdot \prod_{\forall c'\in(\mathcal{C}\backslash c)}\hat{\varepsilon}_{\theta,c,c'}^{DAVID-Q}\label{eq:admission_david_q_eligibility}\\  
\sigma_{\theta,c}^{DAVID-Q}(P^{sub})&=\textbf{1}\big(v_{\theta,c}>\gamma_{\mathcal{I}}\big) \cdot \textbf{1}\big(e_{\theta,c}<P^{r}_c\big)\cdot \textbf{1}\big(w_{\theta,c}>P^{d}_c\big)\cdot \prod_{\forall c'\in(\mathcal{C}\backslash c)}\hat{\sigma}_{\theta,c,c'}^{DAVID-Q}\label{eq:admission_david_q_disclosure}\\
\hat{\varepsilon}_{\theta,c,c'}^{DAVID-Q}(P^{sub})&=
\begin{cases}
1 & \text{if either i) }v_{\theta,c}>v_{\theta,c'}\text{ or ii) }\big(e_{\theta,c'}<P_{c'}^r\text{ and }w_{\theta,c'}<P_{c'}^d\big)\\
 & \text{or iii) }\big(e_{\theta,c'}<P_{c'}^r\text{ and }v_{\theta,c}>v_{\theta,c'}-\gamma_{\mathcal{I}}\big)\\
0 & \text{otherwise}
\end{cases}\label{eq:admission_david_q_eligibility_outside}\\
\hat{\sigma}_{\theta,c,c'}^{DAVID-Q}(P^{sub})&=
\begin{cases}
1 & \text{if either i) }v_{\theta,c}-\gamma_{\mathcal{I}}>v_{\theta,c'}\text{ or ii) }\big(e_{\theta,c'}<P_{c'}^r\text{ and }w_{\theta,c'}<P_{c'}^d\big)\\
 & \text{or iii) }\big(e_{\theta,c'}<P_{c'}^r\text{ and }v_{\theta,c}>v_{\theta,c'}\big)\\
0 & \text{otherwise}
\end{cases}\label{eq:admission_david_q_disclosure_outside}
\end{align}

\begin{align}
\varepsilon_{\theta,c}^{DAVID-U}(\Pi)&=\textbf{1}\big(v_{\theta,c}>0\big) \cdot \textbf{1}\big(E[w_{\theta,c}|e_{\theta,c}]>\Pi_c\big)\cdot \prod_{\forall c'\in(\mathcal{C}\backslash c)}\hat{\varepsilon}_{\theta,c,c'}^{DAVID-U}\label{eq:admission_david_u_eligibility}\\  
\sigma_{\theta,c}^{DAVID-U}(\Pi)&=\textbf{1}\big(v_{\theta,c}>\gamma_{\mathcal{I}}\big) \cdot \textbf{1}\big(E[w_{\theta,c}|e_{\theta,c}]<\Pi_c\big)\cdot \textbf{1}\big(w_{\theta,c}>\Pi_c\big)\cdot \prod_{\forall c'\in(\mathcal{C}\backslash c)}\hat{\sigma}_{\theta,c,c'}^{DAVID-U}\label{eq:admission_david_u_disclosure}\\
\hat{\varepsilon}_{\theta,c,c'}^{DAVID-U}(\Pi)&=
\begin{cases}
1 & \text{if either i) }v_{\theta,c}>v_{\theta,c'}\text{ or ii) }\big(E[w_{\theta,c'}|e_{\theta,c'}]<\Pi_{c'}\text{ and }w_{\theta,c'}<\Pi_{c'}\big)\\
 & \text{or iii) }\big(E[w_{\theta,c'}|e_{\theta,c'}]<\Pi_{c'}\text{ and }v_{\theta,c}>v_{\theta,c'}-\gamma_{\mathcal{I}}\big)\\
0 & \text{otherwise}
\end{cases}\label{eq:admission_david_u_eligibility_outside}\\
\hat{\sigma}_{\theta,c,c'}^{DAVID-U}(\Pi)&=
\begin{cases}
1 & \text{if either i) }v_{\theta,c}-\gamma_{\mathcal{I}}>v_{\theta,c'}\text{ or ii) }\big(E[w_{\theta,c'}|e_{\theta,c'}]<\Pi_{c'}\text{ and }w_{\theta,c'}<\Pi_{c'}\big)\\
 & \text{or iii) }\big(E[w_{\theta,c'}|e_{\theta,c'}]<\Pi_{c'}\text{ and }v_{\theta,c}>v_{\theta,c'}\big)\\
0 & \text{otherwise}
\end{cases}\label{eq:admission_david_u_disclosure_outside}
\end{align}

\begin{figure}[b!]
    \centering
    \begin{subfigure}{.5\linewidth}
        \centering
        \begin{tikzpicture}
    \draw[thick,->] (0,0) -- (4.5,0);
    \draw[thick,->] (0,0) -- (0,4.5); 
    \draw (4.85, 0) node {$P_1$};
    \draw (0, 4.85) node {$P_2$};
    \draw (-.3, 0) node {$0$};
    \draw (0, -.3) node {$0$};
    \draw (-.3, 4) node {$1$};
    \draw (4, -.3) node {$1$};
    \draw (-.3, 1) node {$e^\theta_2$};
    \draw (1, -.3) node {$e^\theta_1$};
    \draw[fill=red!50] (0,4) -- (0,1)  -- (1,1) -- (1,4) -- cycle;
    \draw[fill=blue!50] (0,0) --  (0,1) -- (4,1) -- (4,0) -- cycle;
    \draw[fill=white] (1,1) -- (4,1) -- (4,4) -- (1,4) -- cycle;
    \draw (2.75, 2.75) node {$(\emptyset,\emptyset)$};
    \draw (.5, 2.75) node {$(1,\emptyset)$};
    \draw (2.25, .5) node {$(2,\emptyset)$};
\end{tikzpicture}
        \caption{No information disclosure in DA}
    \end{subfigure}%
    \begin{subfigure}{.5\linewidth}
        \centering
        \begin{tikzpicture}
    \draw[thick,->] (0,0) -- (4.5,0);
    \draw[thick,->] (0,0) -- (0,4.5);
    \draw (4.85, 0) node {$P_1$};
    \draw (0, 4.85) node {$P_2$};
    \draw (-.3, 0) node {$0$};
    \draw (0, -.3) node {$0$};
    \draw (-.3, 4) node {$1$};
    \draw (4, -.3) node {$1$};
    \draw (-.3, 1) node {$e^\theta_2$};
    \draw (1, -.3) node {$e^\theta_1$};
    \draw (-.3, 2) node {$w^\theta_2$};
    \draw (2, -.3) node {$w^\theta_1$};
    \draw[fill=blue!20] (1,1) --  (1,2) -- (4,2) -- (4,1)-- cycle;
    \draw[fill=red!20] (1,2) --  (2,2) -- (2,4) -- (1,4)-- cycle;
    \draw[fill=red!50] (0,4) -- (0,1)  -- (1,1) -- (1,4) -- cycle;
    \draw[fill=blue!50] (0,0) -- (0,1) -- (4,1) -- (4,0) -- cycle;
    \draw[fill=white] (2,2) -- (4,2) -- (4,4) -- (2,4) -- cycle;
    \draw (2.75, 2.75) node {$(\emptyset,\emptyset)$};
    \draw (.5, 2.75) node {$(1,\emptyset)$};
    \draw (2.25, .5) node {$(2,\emptyset)$};
    \draw (2.25, 1.5) node {$(2,2)$};
    \draw (1.5, 2.75) node {$(1,1)$};
\end{tikzpicture}
        \caption{Information disclosure through DAVID-U}
    \end{subfigure}%
    \caption{Individual demand and targeting}
    \label{fig:demand_disclosure}
    \floatfoot{Notes: The figures in the two panels depict individual demand and targeting behavior with and without voluntary information disclosure as a function of college cutoffs. The figure is based on a situation with a student of type $\theta$ and two colleges denoted by 1,2. In the figure we assume that college 2 is preferred, i.e., $v_2^\theta>v_1^\theta$, but not net of targeting costs, $v_2^\theta-\gamma_\mathcal{I}<v_1^\theta$ and that the payoff to each college from admitting a student of type $\theta$ exceeds its eligibility score, i.e. $w^\theta_j>e^\theta_j$ for both $j\in\{1,2\}$. The  elements of the tuple within the shaded areas denote which college is demanded, i.e. where the student can expect admission, and which college is targeted. To illustrate this, consider the red shaded area in panel (a) with the tuple (1,$\emptyset$). In this area, the student is accepted at college 1 and she does not target any colleges. In the light-shaded red area in panel (b) with the tuple (1,1), the student is accepted at college 1 and she targets college 1.} 
\end{figure}
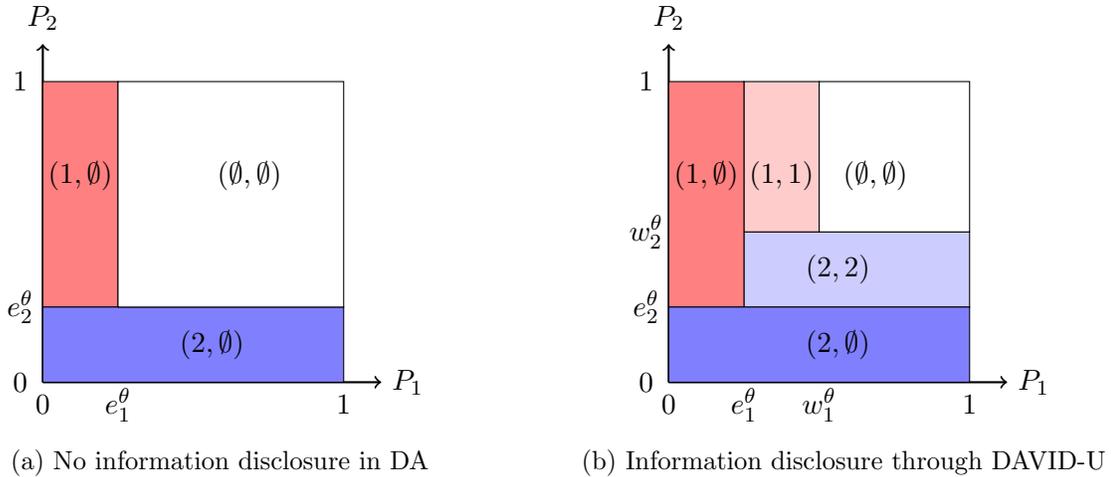

To illustrate that demand in endogenous disclosure satisfies these two properties, we depict an example of demand  with and without voluntary disclosure in Figure~\ref{fig:demand_disclosure}. The figure shows that, irrespective of voluntary information provision, the demand for each college is non-increasing in own cutoff and non-decreasing in the other colleges' cutoff. The lemma below specifies how these properties are generalizable.

\subsection{Supply and demand properties for disclosure without quotas}\label{app:supply_demand_disclosure}

We continue with a separate analysis of the supply and demand framework for DAVID-U as this requires developing further model machinery. 

A central property that \citet{Azevedo2016AMarkets} use to derive results about matchings is that individual demand adjusts to admission cutoffs in the same way that price affects demand in standard consumer choice. One remarkable fact about our matching framework, which extends \citet{Azevedo2016AMarkets} to two kinds of blocking pairs, is that the same fundamental properties of demand about the individuals are preserved. The two properties are defined as follows. We say that individual demand is \textbf{monotone} if $D_{\theta,c}$ is nonincreasing in $\Pi_c$ for any type. Individual demand satisfies \textbf{gross substitutes} if $D_{\theta,c}$ is non-decreasing in $\Pi_{c'}$ for any $c'\ne c$.  


A market clearing cutoff is a vector of cutoffs that clears supply of and demand for colleges.

\begin{definition}
Cutoffs $\Pi$ are market clearing if they satisfy the following market clearing conditions:
\begin{itemize}
    \item $D_c(\Pi)\le S_c$ for all colleges
    \item $D_c(\Pi)= S_c$ for every college where $\Pi_c>-\infty$
\end{itemize}
\end{definition}

\begin{lemma}\label{lemma:demand_grosssub}
At any cutoff it holds that individual demand is monotone and satisfies gross substitutes.
\end{lemma}

\noindent\textbf{Proof for Lemma~\ref{lemma:demand_grosssub}:} We focus on a given college $c$ and a given type $\theta$ and let $\Pi_{-c}$ denote cutoffs of other colleges. We will first prove that demand is \textit{monotone} in $\Pi_c$, i.e. that $D_{\theta,c}(\cdot)$ is nonincreasing in $\Pi_c$. We see from \eqref{eq:admission_david_u_eligibility} and \eqref{eq:admission_david_u_disclosure} that demanding $c$ requires either $E[w_{\theta,c}|e_{\theta,c}]>\Pi_c$ or ($E[w_{\theta,c}|e_{\theta,c}]<\Pi_c$ and $w_{\theta,c}>\Pi_c$). This means that type $\theta$ can only increase demand for college $c$ for a given $\Pi_{-c}$ if $\Pi_c$ is below a given threshold (either $E[w_{\theta,c}|e_{\theta,c}]$ or $w_{\theta,c}$ depending on $\theta$). We will next demonstrate that demand satisfies \textit{gross substitutes}, i.e., $D_{\theta,c}(\cdot)$ is non-decreasing in $\Pi_{c'}$ where $c'\in(\mathcal{C}\backslash c)$. 
We see that type $\theta$ can only stop demanding $c$ (by starting to demand $c'$) if  \eqref{eq:admission_david_u_eligibility_outside} and \eqref{eq:admission_david_u_disclosure_outside} is violated, which requires that $E[w_{c'}^\theta|e_{c'}^\theta]>\Pi_{c'}$ or ($E[w_{c'}^\theta|e_{c'}^\theta]<\Pi_{c'}$ and $w_{\theta,c}>\Pi_c$). In other words, type $\theta$ can only decrease demand for college $c$ keeping $\Pi_{-c'}$ fixed if $\Pi_{c'}$ is below some threshold (either $E[w_{c'}^\theta|e_{c'}^\theta]$ or $w_{c'}^\theta$ depending on $\theta$). $\blacksquare$
\bigskip

An important consequence of the lemma is that all the properties of aggregate demand and matchings carry over from \citet{Azevedo2016AMarkets} to our setting.

\section{Proofs}\label{app:proofs}

\textbf{Proof of Proposition~\ref{proposition:strategyproof}}: 
Prior work has established that DA satisfies ordinal strategy, see  
\cite{Abdulkadiroglu2015ExpandingChoice}.

We proceed with establishing that DAVID-Q satisfies ordinal strategy-proofness by using contradiction. Suppose that ordinal strategy-proofness does not hold for DAVID-Q then there exists a student who is strictly better off by submitting $ROL^{sub}_i$ where the ordering of colleges violates her preferences $v_i$. In other words, there are two colleges $c,c'$ where $v_{i,c}>v_{i,c'}$ but $c'$ is listed ahead of $c$. There exist environments and actions by other students such that $e_{i,c}>P_c$, $e_{i,c'}>P_{c'}$ and for all $c''\in(\mathcal{C}\backslash\{c,c'\})$ such that $v_{i,c''}>v_{i,c}$. However, these situations would imply that student $i$ could increase her utility by re-ranking colleges $c,c'$, which violates the original assumption that the student could benefit by choosing another order. Thus ordinal strategy-proofness must hold.

To establish ordinal strategy-proofness for DAVID-U, we exploit the fact that DAVID-U applies the ordinal rankings submitted by students in the same way as DA. Thus, if for a given profile of student preferences,
targeting behavior and college priorities at which some student could benefit from manipulation of (ordinal) preferences, then the student would benefit from preference manipulation in the corresponding DA that is induced by the same profile (of student preferences and college priorities from eligibility scores only). However, this contradicts strategy-proofness of DA \citep{Roth1982TheIncentives}. Therefore, DAVID-U is strategy-proof with respect to students’ ordinal preferences.
$\blacksquare$\bigskip

\textbf{Proof of Theorem~\ref{theorem:DAVID-Q_safe}}: 
Take an arbitrary truthful partial ordering of the colleges and denote this by $R=[c_1,..,c_R]$. First we focus on DAVID-Q and assume that student $i$ submits $ROL^{sub}_i$ containing regular sub-colleges of colleges in $R$ using the same order. We now investigate what happens if the student includes the information disclosure sub-college of college $c=c_q$ into $ROL^{sub}_i$ and denote the modified ROL as $\tilde{ROL}^{sub}_i$. As  $\tilde{ROL}^{sub}_i$ is transitive and the regular sub-college is ranked ahead of its disclosure sub-college, it follows that there is a unique feasible ROL: $\tilde{ROL}^{sub}_i=[c^{r}_1,..,c^{r}_{j},c^{vid}_{j},c^{r,\theta}_{j+1},..,c^{r}_{R}]$. We see that for any cutoffs, $P$, it holds that the student may only get an improved match in $\tilde{ROL}^{sub}_i$ over $ROL^{sub}_i$. If the student is admitted to colleges $c_1,..,c_q$ through their eligibility score, then the match is the same; otherwise the student may be admitted at $c$ through the information disclosure quota if it has a high enough value, i.e., $w_{i,c}>P_c$. Thus, in any case, the students gets the same or better match quality. Thus we have shown that including an information disclosure sub-college in a ROL that contains only regular sub-colleges cannot lower the match utility. The same property can be extended to any $ROL^{sub}_i$ that already contains information disclosure sub-colleges. 

We now show that the same property does not hold for DAVID-U. Let $ROL_i=R$ and let there be some set of targeted colleges, $x_i\subseteq(\mathcal{C}\backslash c) $. Suppose that $P$ is such that all colleges ranked above $c$ in $ROL_i$ are not affordable and that $E[w_{i,c}|e_{i,c}]\ge P_c$ as well as $w_{i,c}< P_c$. By targeting college $c$,  the student will have lower matching utility, as she will not be admitted to $c$ and instead must either be matched with a less preferred college or be unmatched. Consequently, we have shown that DAVID-U is not safe.
$\blacksquare$\bigskip

\noindent \textbf{Proof of Theorem~\ref{theorem:justifiedenvy_davi}}: 
First, we know from existing work that DA eliminates justified envy \citep{Abdulkadiroglu2003SchoolApproach}.

We proceed to demonstrate the justified envy property for DAVID-Q. By construction, both the regular and disclosure sub-colleges correspond to a college in a standard DA. 
As DA eliminates justified envy, it follows that there will be no justified envy in admission to regular sub-colleges as, here, admission happen according to DA where sub-college priorities equal eligibility scores. Conversely, for disclosure sub-colleges there is no justified envy as admission happens according to DA but where priorities equal the actual payoff (after information disclosure). However, the property does not hold at the college-wide level and across admission criteria; e.g., a student rejected in regular admission may have justified envy of a student admitted after disclosure. 

We assert that the assignment produced by DAVID-U eliminates justified envy with and without disclosure. This entails that there exists a pair of a student and a college that were not matched in DAVID-U where: (i) the student prefers the college to her current assignment; and, (ii) either the expected value conditional on her eligibility score or the college assigned to the student exceeds the minimum expected payoff of admitted students to the college (i.e., the cutoff). However, as the assignment of DAVID-U is produced by DA, it follows that if the student has ranked the college higher than her current assignment, then the student must have been rejected by the college due to a too low expected payoff. $\blacksquare$
\bigskip

\noindent \textbf{Proof of Theorem~\ref{theorem:unique_eqilibria}}: 
The proof proceeds in three steps. First, we verify that the conditions are sufficient for uniqueness of equilibrium in the preference revelation games of DA and DAVID-Q. Second, we also verify that the conditions are sufficient for uniqueness of equilibrium in the preference revelation games of DAVID-U. Third and finally, we show that there exists a disclosure share such that the equilibria of DAVID-Q and DAVID-U coincide.

\textit{Unique equilibrium in DA and DAVID-Q} We begin by establishing the uniqueness of DAVID-Q. 
We note that DAVID-Q generalizes the standard framework where the quota for admission with voluntary information is set to zero, as a consequence, uniqueness also follow for DA. Equilibria at the sub-college level correspond exactly to using a decentralized matching framework where sub-colleges correspond to standard colleges in \citet{Azevedo2016AMarkets}. We can rewrite the sub-college framework such that (i) preferences for regular admission are given by $v_\theta$ and for admission with disclosure, the utility of admission is $v_\theta-\gamma_{\mathcal{I}}$, and (ii) priorities at regular sub-colleges are determined by $e_{\theta}$, and by $w_{\theta}$ at disclosure sub-colleges. This implies that the equilibrium conditions of our preference revelation games for DAVID-Q correspond one-to-one to the stable matching setup of \citet{Azevedo2016AMarkets} and we can apply their results of uniqueness. 

\textit{Unique equilibrium in DAVID-U} 
To establish this part we argue that the essential conditions for establishing the results in \citet{Azevedo2016AMarkets} are fulfilled. 

We begin with extending the results of \citet{Azevedo2016AMarkets}, by removing the limitation of type space where every eligibility score and utility are restricted to $[0,1]$ to instead being restricted to $\mathbb{R}$. This follows as we can define the auxiliary, topological space $\mathbb{R}_{-\infty,\infty}$, which is the union of $\mathbb{R}$ and $\{\infty,-\infty\}$. We can construct a bijective map between this auxiliary space and $[0,1]$, e.g., using the sigmoid function between $\mathbb{R}$ and $(0,1)$ as well as extremum points to each other (i.e. 0,1 in $[0,1]$ and $-\infty,\infty$ in $\mathbb{R}_{-\infty,\infty}$). Next, we observe that the requirements for the type distribution, either having full support or being regular, only put restrictions on the interior of the type space. Therefore, we can let the extremum points have measure zero as they are not part of the interior. We have now shown that the results also hold for types defined on $\mathbb{R}$ rather than only $[0,1]$.

Next, we leverage our Lemma~\ref{lemma:demand_grosssub}, which establishes that the individual demand function does indeed satisfy monotonicity and gross substitutes. Therefore, we may repeat the steps of \citet{Azevedo2016AMarkets} in deriving the same properties as they only require properties of demand. In the following these properties in references to results in \citet{Azevedo2016AMarkets} are listed: First, to show aggregate demand is monotone and satisfies gross-substitutes (Remark A1). Second, to construct an auxiliary map from an initial level of cutoffs to the market clearing cutoff closest with minimal distance to these initial level of cutoffs. \citet{Azevedo2016AMarkets} then uses this map to show existence of stable matchings and that the set of cutoffs is a lattice (respectively Corollary A1 and Theorem A1). Third, use the lattice structure to establish that there are is a smallest and largest cutoff (Proposition A2) as well as establishing a rural hospital theorem (Theorem A2), implying that the measure of students matched to each colleges is the same across equilibria.

To establish the first part of Theorem~\ref{theorem:unique_eqilibria} on uniqueness given that $\eta$ has full support, we note that the proof for Theorem 1, part 1 in \citet{Azevedo2016AMarkets}, does not require anything beyond full support, which can also be assumed in our setting. 

To establish the second part we only note that requirements of regularity put restriction on the aggregate demand for given colleges conditional on the measure $\nu$. However, these properties of aggregate demand either follow directly from results already established or from the assumption of regularity.

\textit{Equivalence matching DAVID-Q and DAVID-U} For each college, let $\delta_c$ equal the share of agents who target college $c$ among those who demand it. For each type, define the modified ROL, let  $\tilde{ROL}^{sub,\theta}=[c^{r,\theta}_1,..,c^{r,\theta}_{l},c^{vid,\theta}_{l},c^{r,\theta}_{l+1},..,c^{r,\theta}_{L}]$ if college $c_{l}$ is targeted by $\theta$ in the unique equilibrium when the mechanism is not quota-based; otherwise $\tilde{L}^\theta=L^\theta$. Suppose that a student type $\theta$ has another matching than the unique equilibrium associated with DAVID-U, then either the student is indifferent about the original college (which has measure zero) or the student must have preferred the originally matched college. Therefore, there cannot be any measurable differences between the matching produced using the quotas chosen and the unique equilibrium associated with DAVID-U. 
$\blacksquare$
\bigskip

\section{Auxiliary tables}\label{app:tables}

\begin{table}[h!]
\begin{tabular}{l*{6}{c}}
\hline\hline
                                   &\multicolumn{1}{c}{(1)}&\multicolumn{1}{c}{(2)}&\multicolumn{1}{c}{(3)}&\multicolumn{1}{c}{(4)}&\multicolumn{1}{c}{(5)}&\multicolumn{1}{c}{(6)}\\
\hline
2nd Quartile Quota 2 Priority Index&      -0.027&      -0.035&      -0.025&      -0.031&      -0.031&      -0.035\\
                                   &     (0.007)&     (0.011)&     (0.011)&     (0.018)&     (0.009)&     (0.014)\\
3rd Quartile Quota 2 Priority Index&      -0.059&      -0.059&      -0.054&      -0.044&      -0.059&      -0.071\\
                                   &     (0.007)&     (0.011)&     (0.011)&     (0.019)&     (0.010)&     (0.015)\\
4th Quartile Quota 2 Priority Index&      -0.081&      -0.091&      -0.075&      -0.084&      -0.078&      -0.096\\
                                   &     (0.007)&     (0.012)&     (0.011)&     (0.019)&     (0.011)&     (0.017)\\
\hline
N                                  &      33,181&      33,181&      15,186&      15,186&      18,003&      18,003\\
Year FE                            &         X&         X&         X&         X&         X&         X\\
Study Program FE                         &         X&         X&         X&         X&         X&         X\\
Eligibility Score FE                           &         X&         X&         X&         X&         X&         X\\
Study program $\times$ Eligibility score FE               &          &         X&          &         X&          &         X\\
Subsample by admission quota                    &         Any&         Any&     Quota 1&     Quota 1&     Quota 2&     Quota 2\\
\hline\hline

\end{tabular}
\caption{Graduation rate and quota priority index}
\label{tab:graduation_rate_priority}
\floatfoot{Notes: Each column shows the estimation result of a linear probability model with graduation as an outcome. Columns (1) and (2) include any admitted student. Columns (3) and (4) only include students admitted through quota 1. Columns (5) and (6) only include students admitted through quota 2. All models include annual fixed effects at year of admission (Year FE). Furthermore, models have fixed effects for each Study program and each Eligibility score at the 0.1 level of precision, e.g., if two students both have an Eligibility score of 6.7. The models in columns (2), (4) and (6) are estimated with an interaction of the fixed effects Study program by Eligibility Score for the interaction between a study program and eligibility score. The regression sample is based on admission in 2012, 2013, and 2014}
\end{table}

\end{document}